\documentclass[fleqn,usenatbib]{mnras}

\usepackage[T1]{fontenc}

\DeclareRobustCommand{\VAN}[3]{#2}
\let\VANthebibliography\thebibliography
\def\thebibliography{\DeclareRobustCommand{\VAN}[3]{##3}\VANthebibliography}

\usepackage{graphicx}	% Including figure files
\usepackage{amsmath}	% Advanced maths commands
\usepackage{amssymb}	% Extra maths symbols
\usepackage{multirow}

\title[QPOs of MAXI J1535]{Wavelet analysis of MAXI J1535-571 with Insight-HXMT}

\author[X. Chen et al.]{
X. Chen,$^{1,2}$\thanks{E-mail: xiao.chen@whu.edu.cn}
W. Wang,$^{1,2}$\thanks{E-mail: wangwei2017@whu.edu.cn}
B. You,$^{1,2}$\thanks{E-mail: youbei@whu.edu.cn}
P. F. Tian,$^{1,2}$
Q. Liu,$^{1,2}$
P. Zhang,$^{1,2}$
Y. Z. Ding,$^{1,2}$
J. L. Qu,$^3$
\newauthor
S. N. Zhang,$^3$
L. M. Song,$^3$
F. J. Lu,$^3$
and S. Zhang$^{3}$
\\
$^{1}$School of Physics and Technology, Wuhan University, Wuhan 430072, China\\
$^{2}$WHU-NAOC Joint Center for Astronomy, Wuhan University, Wuhan 430072, China\\
$^{3}$Key Laboratory of Particle Astrophysics, Institute of High Energy Physics, Chinese Academy of Sciences, Beijing 100049, China
}

\date{Accepted 2022 April 22. Received 2022 April 7; in original form 2021 June 21}

\pubyear{2022}

\begin{document}
\label{firstpage}
\pagerange{\pageref{firstpage}--\pageref{lastpage}}
\maketitle

\begin{abstract}
In this paper, wavelet analysis is used to study spectral-timing properties of MAXI J1535-571 observed by Insight-HXMT. The low-frequency quasi-periodic oscillations (QPOs) are detected in nine observations. Based on wavelet analysis, the time intervals with QPO and non-QPO are isolated separately, and the corresponding spectra with QPO and non-QPO are analyzed. We find that the spectra with QPO (hereafter QPO spectra) are softer than those without QPO (hereafter non-QPO spectra) in the hard intermediate state (HIMS). While in the soft intermediate state (SIMS), the QPO spectra are slightly harder. The disk temperature of QPO regime is slightly lower during HIMS, but becomes higher during SIMS. The cutoff energies of QPO spectra and non-QPO spectra do not show significant differences. The flux ratio of the disk to total flux is higher for the time intervals with non-QPO than that of QPO regime. We propose that these differences in the spectral properties between QPO and non-QPO regimes could be explained in the scenario of Lense-Thirring precession, and the reversal of the QPO/non-QPO behavior between HIMS and SIMS may be associated with appearance/disappearance of a type-B QPO which might origin from the precession of the jet.
\end{abstract}

\begin{keywords}
Black holes physics -- X-rays: binaries -- Stars: individual (MAXI J1535-571)
\end{keywords}

%%%%%%%%%%%%%%%%%%%%%%%%%%%%%%%%%%%%%%%%%%%%%%%%%%

%%%%%%%%%%%%%%%%% BODY OF PAPER %%%%%%%%%%%%%%%%%%

\section{Introduction}
%2006.09077 many ref
Black hole (BH) transients are shortly detected during their outbursts. They are quite worthy study in X-rays because of the interesting variation in the timing and spectral properties\citep{Kara2019,You2021}. Typically, the spectral states of a BH transient in an outburst evolve through hard (HS), hard/soft intermediate (HIMS/SIMS), soft (SS), intermediate states again, and finally back to a hard state\citep{Remillard2006,Belloni2010}.

Low frequency quasi-periodic oscillations (LFQPOs), with their centroid frequencies ranging from 0.01 to 30 Hz, are commonly discovered in stellar mass BHs \citep{Motta2015,Motta2016,Ingram2019,Ma2021}. Three types (A, B, C) of LFQPOs are identified based on their power spectrum features \citep{Remillard2002,Casella2005,Ingram2019,Xiao2019,Zhang2021}. Studies reveal that relations exist between LFQPO types and spectral states: type-A/B QPOs are normally seen in SIMS \citep{Belloni2020,Garcia2021}, while type-C QPOs can be found in HS, HIMS and SS \citep{Nandi2012,Munoz2014,deRuiter2019}.

The origin and properties of LFQPOs are still inconclusive. A number of models are proposed to explain type-C QPOs \cite[e.g.][]{Molteni1996,Stella1999,Tagger1999,Wagoner1999,Kato2001,Schnittman2006,Chakrabarti2008,Ingram2009,Cabanac2010}. Many of these theories are based on the Lense-Thirring precession, and generally can be categorized as geometrical and intrinsic models \citep{You2018,You2020,Ingram2019}. Observations have shown that the generation of type-C QPOs is related to geometry and/or precession \cite[e.g.][]{Ingram2012,Motta2015,Axelsson2016}, but no clear conclusion regarding Lense-Thirring precession is made \citep{Ingram2019}. For type-B and type-A QPOs, the physical origin is even less clear, except that type-B QPOs may be connected to jets \citep{Fender2009,Motta2015,Russell2020}.

In the past decades, the sudden appearance or disappearance of all three types of QPOs have been discovered in several sources \cite[e.g.][]{Miyamoto1991,Nespoli2003,Huang2018,Xu2019,Sriram2021}. These fast transitions are sometimes associated with type variations, mostly type-A/B transition \citep{Nespoli2003,Sriram2012,Sriram2013,Sriram2016}, but type-B/C \citep{Homan2020} or even type-C/A \citep{Bogensberger2020} transitions are also reported. Many of these QPO timing studies are based on dynamical power density spectra (PDS) technique, with a few tens of seconds intervals \cite[e.g.][]{Zhang2021,Sriram2021}, then have a limit of the time resolution at tens of seconds. Wavelet analysis, on the other hand, can provide accurate time-frequency space information with sufficient small time intervals, thus can be used to study the detailed variations of the periodic or quasi-periodic signals over time \citep{Ding2021}.

The outburst of the Galactic BH transient MAXI J1535-571 was first detected during September 2017, simultaneously by MAXI/GSC \citep{Negoro2017} and Swift/BAT \citep{Kennea2017}, and followed by radio \citep{Russell2017,Tetarenko2017}, sub-millimeter \citep{Tetarenko2017}, near-infrared \citep{Dincer2017} and optical \citep{Scaringi2017} observations. The state transitions \citep{Nakahira2017,Palmer2017,Shidatsu2017,Tao2018,Russell2019,Cuneo2020} and LFQPOs \citep{Gendreau2017, Mereminskiy2017,Bhargava2019,Chatterjee2020,Vincentelli2021} of this source were discussed. Insight-HXMT observation was also performed on September 2017, and its state transitions and QPOs were studied \citep{Huang2018,Kong2020}.

In this paper, the wavelet analysis is utilized to study the 2017 outburst of MAXI J1535-571 with Insight-HXMT data. The observations and data reductions are introduced in Section 2. Section 3 gives the details of data analysis, including the wavelet methods, data separation with QPOs and non-QPOs, and spectral analysis. In Section 4, the distinctions between QPO spectra and non-QPO spectra are discussed. Finally in Section 5, we summarize the results and the implications are presented.

%All papers should start with an Introduction section, which sets the work
%in context, cites relevant earlier studies in the field by \cite{Fournier1901},
%and describes the problem the authors aim to solve \cite[e.g.][]{vanDijk1902}.
%Multiple citations can be joined in a simple way like \cite{deLaguarde1903, delaGuarde1904}.

\section{Observations and data reduction}

Insight-HXMT, the first Chinese X-ray satellite, covers a broad energy band. It contains three telescopes with different energy ranges: the High Energy X-ray telescope (HE) has 18 NaI/CsI detectors from 20-250 keV with $\sim$ 5100 $cm^2$ geometrical area, the Medium Energy X-ray telescope (ME) covers 5-30 keV by 1728 Si-PIN detectors with collecting area of 952 $cm^2$, while swept charge device is used in the Low Energy X-ray telescope (LE) covering 1-15 keV energy range with a total collecting area of 384 $cm^2$ \citep{Zhang2020}. The typical Field of Views (FoVs) are $1.1\degr\times5.7\degr$, $1\degr\times4\degr$ and $1.6\degr\times6\degr$ for HE, ME and LE, respectively \citep{Zhang2020}.

The MAXI J1535-571 observations were performed by Insight-HXMT between September 6 to 18 in 2017. A gap is noticed in the downloaded data between September 07 to 12 due to the X9.3 solar flare. These observation details are discussed in \cite{Kong2020} and \cite{Huang2018}. After the outburst, ten more observations were triggered sporadically in February 2018, and are also included in our data processing.

The Insight-HXMT Data Analysis Software (HXMTDAS) v2.04 is used for data analyzing. The pointing offset angle is smaller than 0.04$\degr$; the elevation angle is greater than 10$\degr$; the geomagnetic cutoff rigidity is larger than 8$\degr$; data within 300 s of the South Atlantic Anomaly (SAA) passage are not used. The light curves are made by HXMTDAS tasks helcgen, melcgen and lelcgen with 0.01 sec time bins. The official tools HEBKGMAP, MEBKGMAP and LEBKGMAP of version 2.0.12 are adopted for both the spectrum and light curve background estimation.

%\begin{equation}
%    x=\frac{-b\pm\sqrt{b^2-4ac}}{2a}.
%	\label{eq:quadratic}
%\end{equation}
%
%Refer back to them as e.g. equation~(\ref{eq:quadratic}).

\section{Methods}
\subsection{Wavelet analysis}

With the time-frequency space information provided by wavelet analysis, the variance of power with time and frequency can be easily identified, so as to distinguish the time periods with QPOs and non-QPOs. \cite{Torrence1998} introduced wavelet analysis methods with elaborated step-by-step guide and examples. In short, a discrete Fourier transform of the time series (i.e. the light curves generated by HXMTDAS tasks subtracted by the corresponding background light curves) is first performed. Then a wavelet function, which is Morlet in our case, is chosen, and has the following form:
\begin{equation}
    \Psi_0(\eta)=\pi^{-1/4}e^{im\eta}e^{-\eta^2/2},
\end{equation}
where $m$ is the nondimensional frequency, $\eta$ is a nondimensional time parameter, and $\Psi_0$ means the $\Psi$ has not been normalized. The reason of performing normalization to the wavelet function is to ensure that the results are comparable to other scales and time series. Thus for each wavelet scale $s$, the normalized wavelet function has unit energy:
\begin{equation}
    \hat{\Psi}(s\omega_k)=(\frac{2\pi s}{\delta t})^{(1/2)} \hat{\Psi}_0(s\omega_k),
\end{equation}
where $\omega_k$ is the angular frequency (see Equation 5 of \cite{Torrence1998} for the definition). Suppose one has N points, then the following relation should be satisfied: 
\begin{equation}
    \sum_{k=0}^{N-1} |\hat{\Psi}(s\omega_k)|^2 = N.
\end{equation}
Finally based on the convolution theorem, the wavelet transform at that scale is the inverse Fourier transform of
\begin{equation}
    W_n(s)=\sum_{k=0}^{N-1} \hat{x}_k \hat{\Psi}^*(s\omega_k) e^{i \omega_k n \delta t}.
\end{equation}
Here the $*$ means the complex conjugate, and $\hat{x}_k$ is the discrete Fourier transform of the time series $x_n$. As the time index, the frequency index and the wavelet scale change, a diffuse two-dimensional time-frequency graph is eventually displayed. A more detailed introduction and equations can be found in \cite{Torrence1998} and \cite{Ding2021}. Since the wavelet result $W_n(s)$ is a complex number because of the wavelet function, we define the wavelet power as $|W_n(s)|^2$.

Different 'mother' wavelet and wavelet parameter $m$ will affect the final results. As indicated by \cite{DeMoortel2004}, the Morlet wavelet or a larger value of $m$ makes the frequency resolution better, while the Paul wavelet or a smaller value of $m$ gives a better time resolution. Since both frequency resolution and time resolution are important in our study, Morlet wavelet with $m=6$ has been taken to achieve a compromise, and to satisfy the admissibility condition \citep{Farge1992}. As a comparison, Paul wavelet with $m=6$ and dynamical PDS with time resolution of 1 s are also tested, and both provide similar peak numbers and locations around the QPO frequency (see Appendix A for a comparison of dynamical power spectrum and wavelet analysis). Paul wavelet however, performs poorly in frequency resolution, and the confidence areas in time are quite narrow, which may affect the fitting results of the QPO spectra. While for the dynamical PDS, the windowed Fourier transform utilized is inaccurate and inefficient \citep{Kaiser2011}.

To increase significance levels, the global wavelet spectrum (i.e. the time-averaged wavelet spectrum) is always performed:
\begin{equation}
    \overline{W}^2(s)=\frac{1}{N} \sum_{n=0}^{N-1} |W_n(s)|^2.
\end{equation}
As discussed in \cite{Torrence1998}, the local wavelet power spectrum and the Fourier spectrum are identical on average, but the global wavelet spectrum provides a better unbiased and consistent estimation \citep{Percival1995}.

The determination of significance levels was also discussed in \cite{Torrence1998}. A white or red noise background spectrum should be appropriately chosen, and the actual spectrum can be compared with this random distributed background. In our case, the lag-1 autoregressive [AR(1)] is used for red noise calculation. At each point $(n,s)$ in the local wavelet power spectrum, the degree of freedom is two. Based on the chi-squared distribution, if a power in the wavelet power spectrum is above the 95\% confidence level (i.e. significant at the 5\% level) compared with the background spectrum, then it can be considered as a true signal.

Since smooth, continuous variations in time series can improve accuracy, and giant time intervals exist in nearly all light curves because of the good time intervals (GTI), thus we separate the light curve data if the adjacent time difference is greater than 0.1 s, which is ten times of the time binsize. As a consequence, HE and ME light curves are normally split into $\sim$ 5 time ranges, while LE contains 1 or 2 separated data, basically the same as the GTI segments (see Appendix B for the detailed GTI information of three instruments). 

\begin{figure*}
	\includegraphics[width=0.95\textwidth]{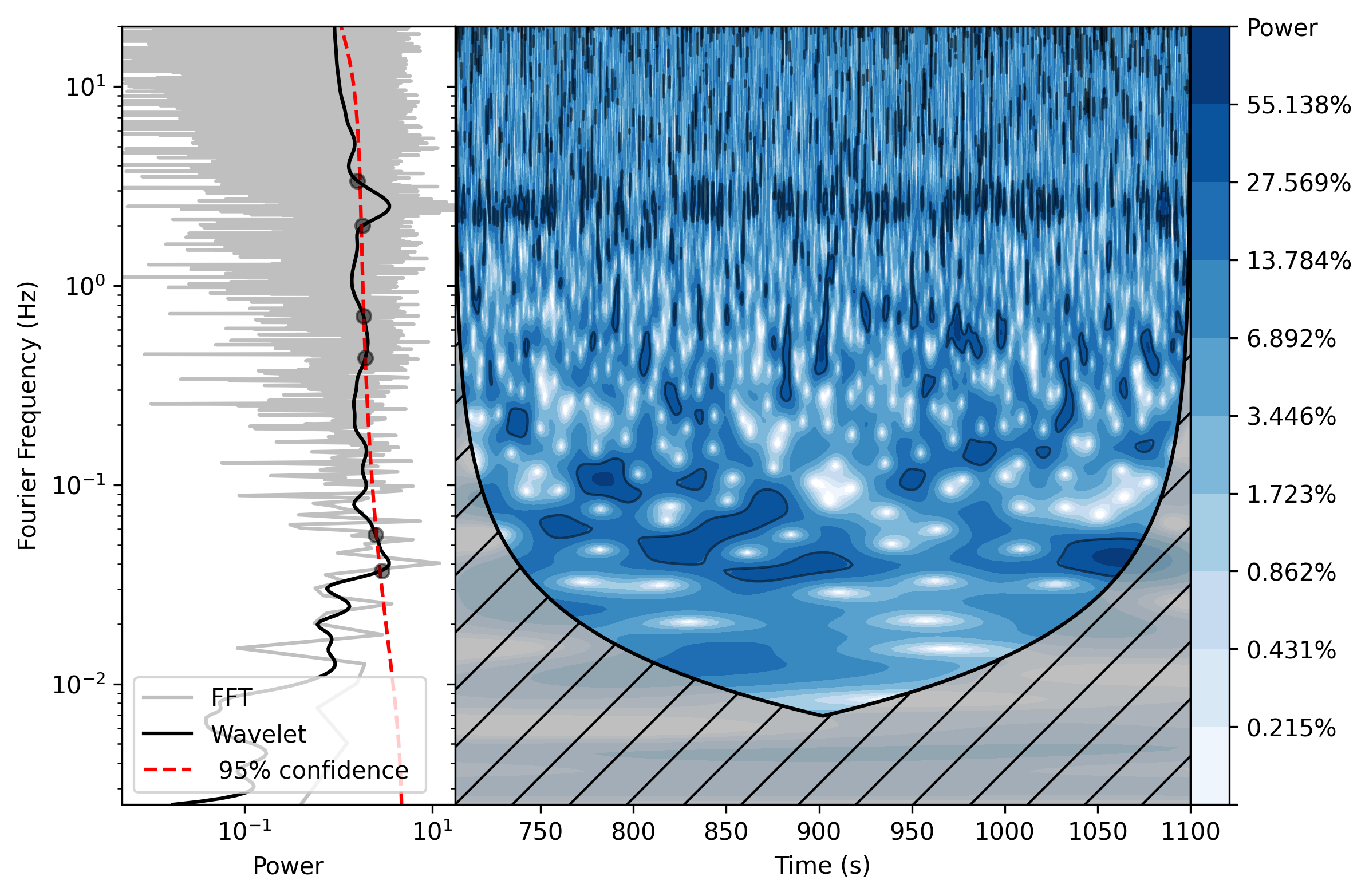}
	\caption{Global wavelet spectra (left, black line) and contour plot (right) of wavelet results for Obs 144. In the left window, PDS result is shown in grey line, 95\% confidence level is plotted with red dash line, and the cross points of global wavelet spectra and 95\% confidence level are presented with dark circles. In the contour plot, the black line circled areas are signals that exceed the 95\% confidence level, and the gray hashed area refer to the cone of influence.}
	\label{fig:example_wa}
\end{figure*}

Figure~\ref{fig:example_wa} gives an example plot of the wavelet analysis. Time series used here is one of the separated HE light curve of observation P011453500144 (hereafter the observation ID will be written as the last three digits for short, i.e. Obs 144) with background subtracted. The global wavelet power spectrum (black line) is presented on the left plot, along with the PDS results (gray line) for comparison. The 95\% confidence level for the global wavelet spectrum is also presented with red dash line. The local wavelet power spectrum is shown on the right side. As power rises up, the plot color changes from white to dark blue. Regions above the 95\% confidence level are enclosed with black contour. Cross-hatched regions at the bottom indicate the cone of influence area, which is caused by the temporarily padding (with zeros) at the end of the time series before performing wavelet transform. The padding enables edge time series to be calculated with wavelet analysis, but also reduces the credibility of that time series. Thus we weighted this area by a factor of 0.1 when plotting. In this figure, a peak can be well noticed in PDS, global wavelet spectrum and local wavelet powers. This demonstrates the reliability of the wavelet algorithm and also provides a basis for the next step.

The wavelet results are discrete data, the step of the logarithmic period is 0.015, which gives a step size of $\sim$ 0.3 Hz around 9 Hz and $\sim$ 0.085 Hz around 2.5 Hz, thus shall provide a relatively reliable data. Details regarding data separation are discussed in the next subsection. The differences of the peaks between GTIs of the same observation is normally less than 0.3 Hz. Here we weight the global wavelet spectra by time and sum them together for each observation, then the QPO information and the mean count rate of each observation are shown in Table~\ref{tab:QPO} for comparison with the previous studies. Since the advantage of wavelet analysis is to provide the detailed time-frequency information, and the global wavelet spectrum is not as efficient as PDS in power spectrum analysis \citep{Liu2007,Bravo2014}, only the central frequency and its full width at half maximum (FWHM) are provided. The QPO centroid frequency increases gradually from $\sim$ 2.5 Hz to $\sim$ 9 Hz, except Obs 301. \cite{Kong2020} and \cite{Huang2018} reported the QPO centroid frequencies derived from the energy range of 27.4 - 31.2 keV and 6 - 38 keV respectively. Compared with our ME results got from 10 - 27 keV, the frequencies are consistent. Our LE (2-10 keV) QPO centroid frequencies are basically the same as the NICER (0.2-10 keV) and Swift/XRT (0.3 – 10 keV) results both reported by \cite{Stiele2018}, i.e., the QPO frequency firstly dropped from $\sim$ 2.5 Hz to 2 Hz, then rose to 3 Hz at around Obs. 601, and finally rose to $\sim$ 9 Hz at Obs. 901.

\begin{table*}
    \scriptsize
	\centering
	\caption{The QPO centroid frequency, FWHM, and mean count rate of each observation. The observation IDs are the last three digits of P011453500XXX. }
	\label{tab:QPO}
	\begin{tabular}{cccccccccccc}
		\hline
		Obs & Start Time (s) &  \multicolumn{3}{|c|}{LE} & \multicolumn{3}{|c|}{ME} & \multicolumn{3}{|c|}{HE}  & state \\
		   & & QPO $\upsilon$ (Hz) & FWHM & count rate & QPO $\upsilon$ (Hz)& FWHM & count rate & QPO $\upsilon$ (Hz) & FWHM  & count rate & \\
		\hline
144	&2017-09-12T10:38:15	&	$	2.61 	\pm	0.01 	$	&	0.79 	&	1168.09 	&	$	2.59 	\pm	0.01 	$	&	0.83 	&	358.52 	&	$	2.56 	\pm	0.01 	$	&	0.76 	&	533.00 	&	HIMS	\\
145	&2017-09-12T13:58:12	&	$	2.56 	\pm	0.01 	$	&	0.87 	&	1188.39 	&	$	2.61 	\pm	0.01 	$	&	0.90 	&	364.35 	&	$	2.61 	\pm	0.01 	$	&	0.87 	&	548.24 	&	HIMS	\\
301	&2017-09-15T04:48:00	&	$	2.14 	\pm	0.01 	$	&	0.66 	&	1309.80 	&	$	2.09 	\pm	0.01 	$	&	0.67 	&	448.36 	&	$	2.02 	\pm	0.01 	$	&	0.62 	&	684.73 	&	HIMS	\\
401	&2017-09-16T06:15:29	&	$	2.79 	\pm	0.01 	$	&	0.84 	&	1482.88 	&	$	2.79 	\pm	0.01 	$	&	0.87 	&	440.11 	&	$	2.79 	\pm	0.01 	$	&	0.84 	&	607.65 	&	HIMS	\\
501	&2017-09-17T06:07:38	&	$	3.36 	\pm	0.01 	$	&	1.01 	&	1713.80 	&	$	3.40 	\pm	0.01 	$	&	1.08 	&	449.40 	&	$	3.40 	\pm	0.01 	$	&	1.06 	&	576.74 	&	HIMS	\\
601	&2017-09-18T02:48:54	&	$	3.32 	\pm	0.01 	$	&	1.02 	&	1759.82 	&	$	3.38 	\pm	0.01 	$	&	1.07 	&	469.73 	&	$	3.35 	\pm	0.01 	$	&	1.07 	&	615.66 	&	HIMS	\\
901	&2017-09-21T02:26:26	&	$	8.59 	\pm	0.04 	$	&	2.42 	&	2675.21 	&	$	9.07 	\pm	0.01 	$	&	2.85 	&	361.92 	&	$	9.28 	\pm	0.02 	$	&	2.88 	&	371.80 	&	SIMS	\\
902	&2017-09-21T06:00:41	&	$	8.45 	\pm	0.16 	$	&	2.71 	&	2685.83 	&	$	9.22 	\pm	0.02 	$	&	2.89 	&	351.46 	&	$	9.32 	\pm	0.02 	$	&	2.91 	&	375.27 	&	SIMS	\\
903	&2017-09-21T09:21:07	&	$	7.59 	\pm	0.08 	$	&	2.58 	&	2626.91 	&	$	8.21 	\pm	0.01 	$	&	2.97 	&	384.63 	&	$	9.07 	\pm	0.03 	$	&	2.90 	&	405.24 	&	SIMS	\\

		\hline
	\end{tabular}
\end{table*}

\subsection{Time split \& spectral analysis}

Based on the wavelet results, we separate data by the 95\% confidence level. Since the peak area on the local wavelet spectrum of Figure~\ref{fig:example_wa} is quite dense because of the long time range, a $\sim$ 40 s time segment result is shown in Figure~\ref{fig:example_wa_amplify}, to elaborate details on the selection of the time period. Once the peak frequency of the global wavelet power curve is confirmed, the 95\% confidence interval of the frequency can be chosen, i.e. the intersection range between the global wavelet power curve and the 95\% confidence red line near the peak. Then if any point in the local wavelet plot within this frequency interval is greater than 95\% confidence level (the points within the black contour enclosed area in Figure~\ref{fig:example_wa_amplify}), then the point is selected as QPO time range. Using this method, we select out all the QPO time ranges in each observation, then $hxmtscreen$ tool in HXMTDAS is used to create a new FITS file that only contains these segments, to replace the original screened event FITS file. Then normal HXMTDAS spectrum generation pipeline steps are executed to generate a new spectrum that just contains the QPO time segment information. All the other remaining time ranges are below 95\% confidence level. Mathematically speaking, they can not be guaranteed as QPO excluded segments. However, they will be called the non-QPO segments from now on just for convenience. Similarly, the non-QPO time ranges are used for non-QPO spectrum generation. Finally for each observation, we have three spectra: one for time-averaged, one for QPO included and one for QPO excluded.

\begin{figure}
	\includegraphics[width=\columnwidth]{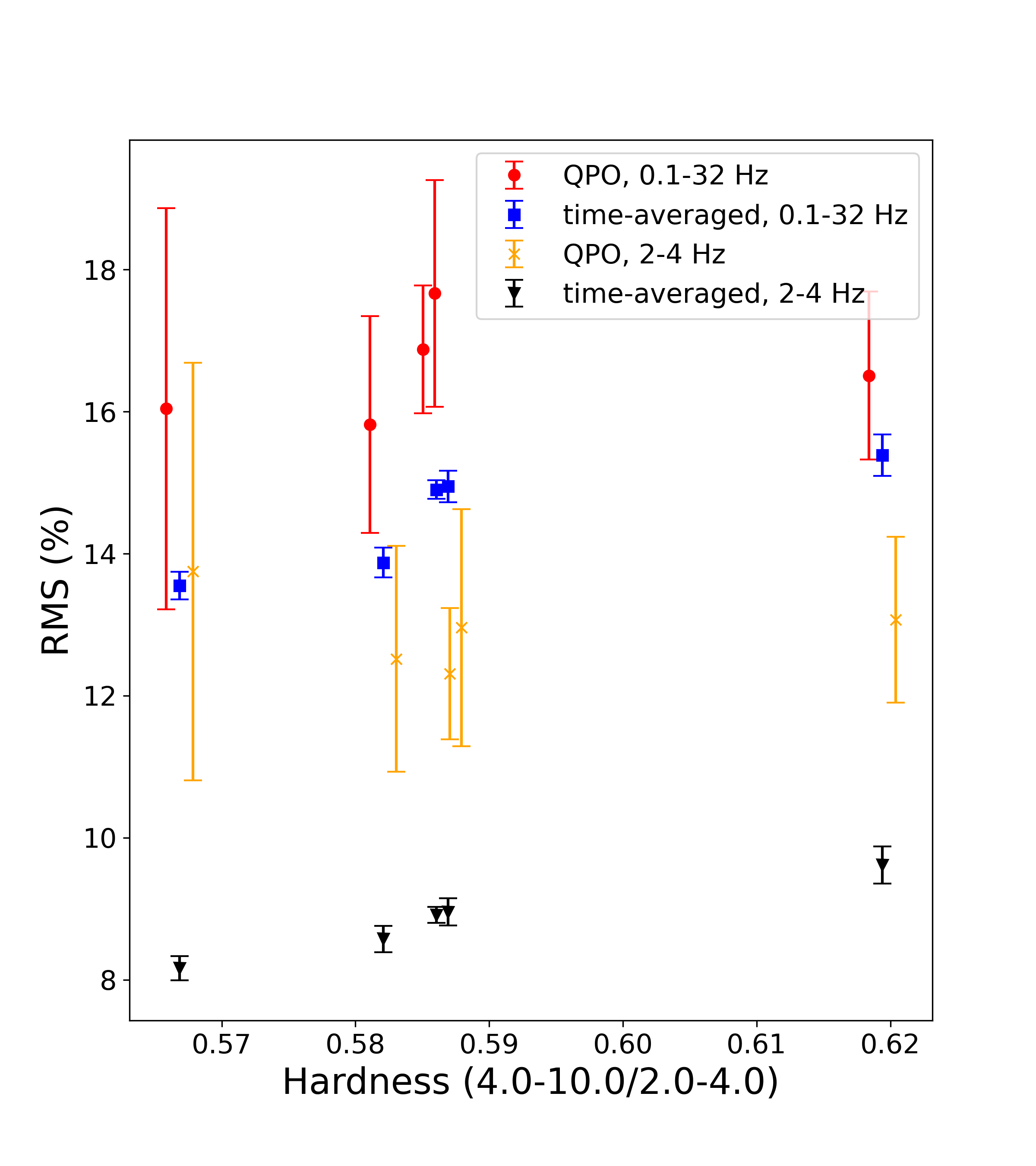}
	\caption{Hardness-rms diagram, where hardness is the ratio of mean count rates of 4.0-12.0 keV to 2.0-4.0 keV. The red circles, blue squares, orange crosses and black inverted triangles represent rms values in the QPO regime in 0.1-32 Hz, the time-averaged regime in 0.1-32 Hz, the QPO regime in 2-4 Hz, and the time-averaged regime in 2-4 Hz, respectively. Only Obs 144-501 are plotted for both QPO and time-averaged regimes. The QPO regimes have large error bars, and are shifted a little to the left (0.1-32 Hz) and right (2-4 Hz) to avoid overlapping.}
	\label{fig:rms-QPO}
\end{figure}

Before further considering the separated spectrum, we first make an analysis of the separated time segments, and the rms results are shown in Figure~\ref{fig:rms-QPO}, including the rms of QPO segments in 0.1-32 Hz and 2-4 Hz, and the rms of the whole time segments in 0.1-32 Hz and 2-4 Hz. Since the QPO time ranges are quite narrow, typically from about a second to a few seconds, a 2-s data interval is chosen for PDS production for both QPO segments and the whole time, to ensure consistency. As a consequence, rms values in the QPO regime and time-averaged regime are obtained for Obs 144-501, and the rms values versus hardness are shown in Figure~\ref{fig:rms-QPO}. Obviously, the QPO regime has larger error bars, since the number of averaged time segments is quite smaller compared to the time-averaged regime. A positive correlation between the rms values and hardness can be noted in the time-averaged regime (e.g., Spearman rank correlation $r=0.956$ and $p=0.01$ for 2-4 Hz). However, in the case of QPO regime, this relationship is not clear. If this change is not caused by error bars, then the positive correlation is not related to the appearance of QPO. The rms value during the QPO regime is much larger than the time-averaged scenario, both for 0.1-32 Hz and 2-4 Hz. The averaged rms of the five points in the QPO regime raised $\sim 15\%$ in 0.1-32 Hz, and $ \sim 50\%$ in 2-4 Hz, compared to the time-averaged regime. We also calculate the rms values in 0.1-2 Hz and 4-32 Hz, and find that they almost remain unchanged between the two regimes (changing by $\sim -6\%$ and 0.2\% for 0.1-2 Hz and 4-32 Hz, respectively). The above results may indicate that, first of all, the light curves in the QPO regime are indeed different from the time-averaged data. Secondly, this change is mainly concentrated around the QPO frequency, rather than in other frequency bands.

\begin{figure}
    \begin{minipage}{0.5\textwidth}
		\includegraphics[width=1\textwidth]{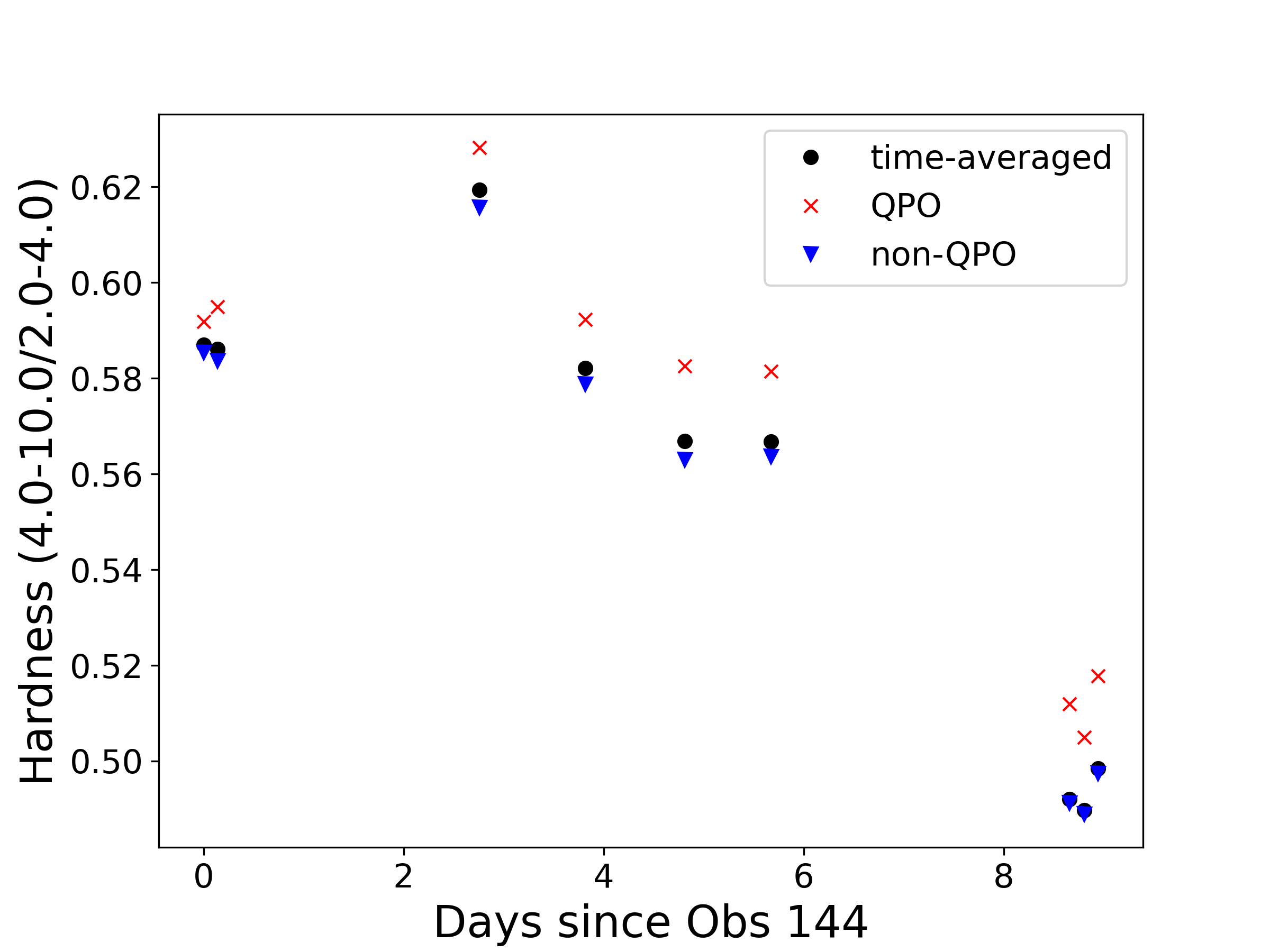}
	\end{minipage}
	\begin{minipage}{0.5\textwidth}
		\includegraphics[width=1\textwidth]{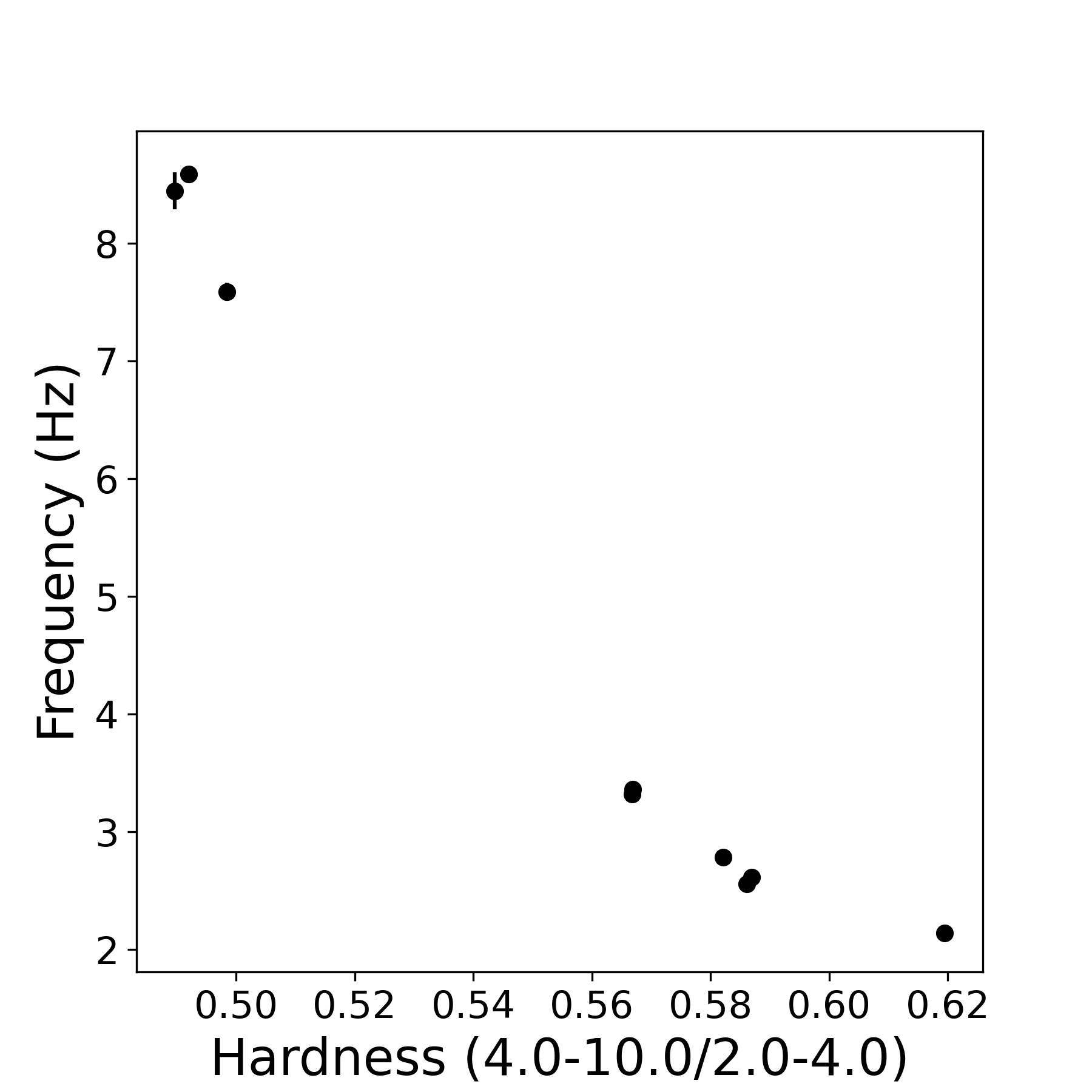}
	\end{minipage}
	\caption{{\bf Top panel:} The hardness evolution with time, hardness is defined as the ratio of mean count rates of 4.0-12.0 keV to 2.0-4.0 keV, and the horizontal axis is the time since Obs 144 in days. Hardness in time-averaged (black circle), QPO (red cross) and non-QPO (blue triangle) regimes are plotted. {\bf Bottom panel:} Centroid QPO frequency-hardness diagram, hardness has the same definition as in the top panel, except that only time-averaged results are shown, and here frequency is the centroid QPO frequency of LE light curves.}
	\label{fig:time-HR-QPO}
\end{figure}

\begin{figure*}
	\includegraphics[width=0.95\textwidth]{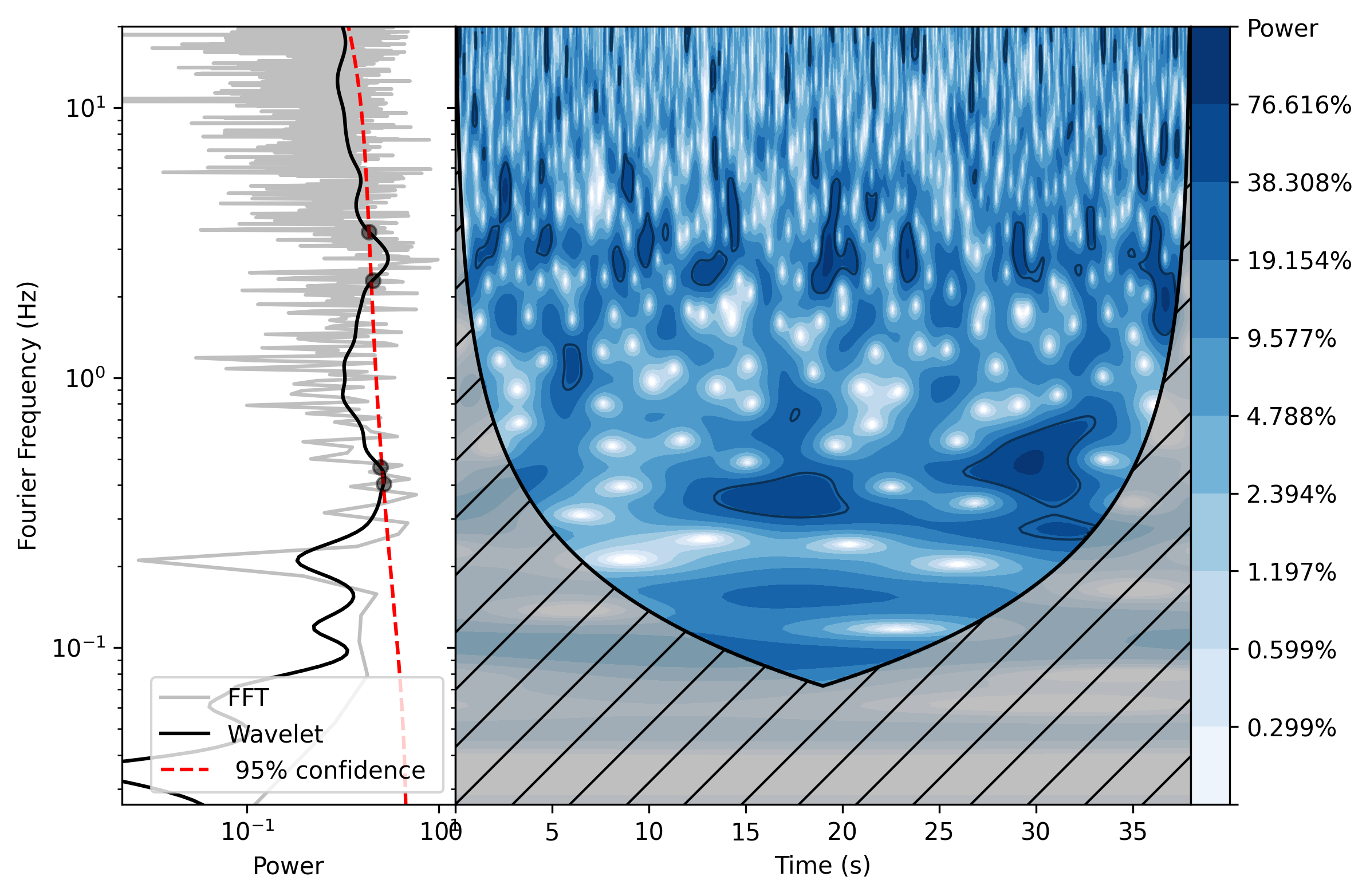}
	\caption{Time selection diagram for a $\sim 40 s$ wavelet result. Elements in the plot are basically the same as Figure~\ref{fig:example_wa}.}
	\label{fig:example_wa_amplify}
\end{figure*}

Figure~\ref{fig:time-HR-QPO} also gives the evolution of hardness (4.0-12.0 keV to 2.0-4.0 keV) with time (top panel), and the QPO frequency-hardness diagram (bottom panel). The hardness basically shows a decreasing trend with time, except Obs 301, which also shows a different trend in the evolution of QPO frequency (see Table~\ref{tab:QPO}). In addition, the diagram of QPO frequency and hardness shows an anti-correlation with the Spearman rank correlation $r=-0.881$ and $p=0.001$. The top panel of Figure~\ref{fig:time-HR-QPO} also indicates that the difference exists between QPO regime and non-QPO regime, and the hardness of the QPO regime is larger than the non-QPO one.

\begin{figure*}
	\begin{minipage}{0.95\textwidth}
		\includegraphics[width=1\textwidth]{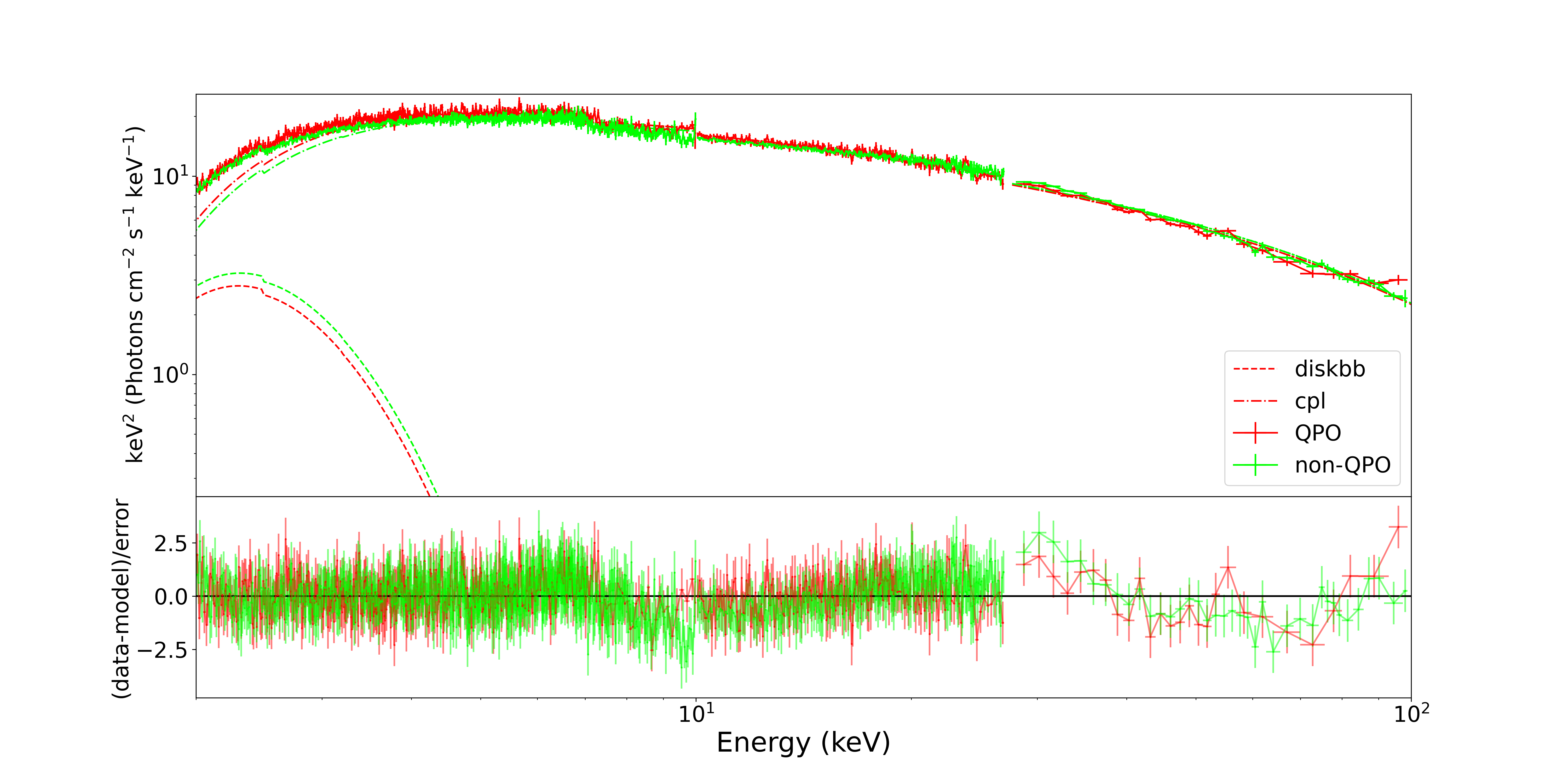}
	\end{minipage}
	\begin{minipage}{0.95\textwidth}
		\includegraphics[width=1\textwidth]{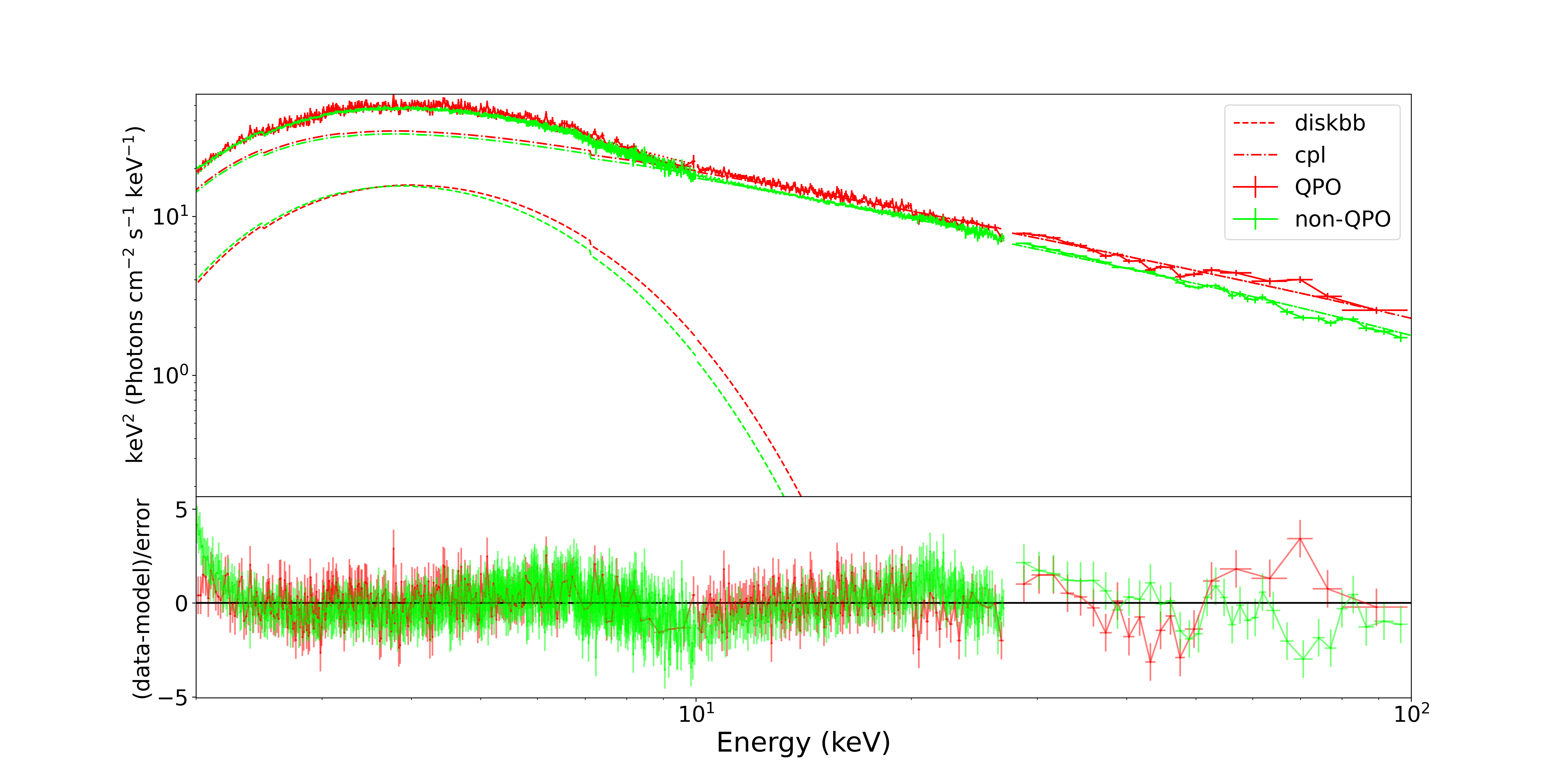}
	\end{minipage}
	\caption{A combo plot of QPO (red) and non-QPO (green) fitting spectra for Obs 144 (top) and 901 (bottom). The energy band is from 2--100 keV, with 2--10 keV for LE, 10--27 keV for ME and 27--100 keV for HE. Model: const*tbabs*(diskbb+cutoffpl), with $N_H$ fixed to $3.8\times 10^{22} \rm \ cm^{-2}$.}
	\label{fig:example_3combo}
\end{figure*}

Next, we use the XSPEC v12.12.0 software package \citep{Arnaud1996} to fit these spectra. The spectra are grouped to have at least 30 counts in each energy bin. \cite{Xu2018} reported a strong iron line in this source in the Hard State (HS), which was also reported with the Insight-HXMT in the hard state \citep{Kong2020}. However, when the source entered the HIMS and SIMS, this Fe $K_{\alpha}$ line was hardly seen (also see \citealt{Kong2020}). Thus similar to the previous work, we consider const*tbabs*(diskbb+cutoffpl) as the model in the spectral fitting, and the results are shown in Table~\ref{tab:xspec}. During the SIMS state (i.e. Obs 901--903), the values of $E_{\rm cut}$ are much higher than 100 keV with large uncertainties, resulting in no physical meaning, so we fix them at 400 keV. Figure~\ref{fig:example_3combo} presents the examples of the spectral fitting results. The top panel shows spectra of Obs 144 in HIMS, while the bottom panel shows those of Obs 901 in SIMS. A systematic error of 2\% for LE/ME/HE was added. In this study, the uncertainty of the best-fitting parameters corresponds to the 90\% confidence level. The energy bands adopted for analysis are 2--10 keV (LE), 10--27 keV (ME) and 27--100 keV (HE).

\begin{table*}
	\centering
	\caption{Spectral fitting parameters. The model used here is const*tbabs*(diskbb+cutoffpl), with $N_H$ fixed to $3.8\times 10^{22} \rm \ cm^{-2}$. A systematic error of 2\% is adopted when calculating $\chi^2$. The exposure time after data reduction is listed at the last column.}
	\label{tab:xspec}
%	\begin{threeparttable}
	\begin{tabular}{cccccccc}
		\hline
		Obs	&	 Tin	 &	 diskbb norm	 &	 $\Gamma$	 &	 $E_{cut}$	 &	 cutoffpl norm	 &	 $\chi^2/d.o.f.$ & exposure \\
		     & (keV) & ($10^5$) & & (keV) & & & (s)
		 \\
		      & &  & &  & & & LE/ME/HE
		 \\
		\hline
		%\tnote{1}
		\multicolumn{8}{|c|}{\bf{time-averaged}} \\
		144	&$	0.38 	\pm	0.01 	$&$	2.56 	_{-	0.54 	}^{+	0.73 	}$&$	2.27 	\pm	0.01 	$&$	68.83 	_{-	2.79 	}^{+	3.02 	}$&$	37.60 	_{-	0.80 	}^{+	0.80 	}$&	1160.32 	/	1275	&	1137	/	2168	/	1684	\\
145	&$	0.37 	\pm	0.01 	$&$	3.08 	_{-	0.56 	}^{+	0.73 	}$&$	2.29 	\pm	0.01 	$&$	69.80 	_{-	1.85 	}^{+	1.94 	}$&$	39.63 	_{-	0.59 	}^{+	0.59 	}$&	1159.93 	/	1275	&	3181	/	8202	/	9329	\\
301	&$	0.34 	\pm	0.01 	$&$	5.89 	_{-	1.42 	}^{+	2.00 	}$&$	2.16 	\pm	0.01 	$&$	53.12 	_{-	1.63 	}^{+	1.71 	}$&$	37.49 	_{-	0.75 	}^{+	0.75 	}$&	1163.08 	/	1275	&	599	/	3380	/	1636	\\
401	&$	0.35 	\pm	0.01 	$&$	4.51 	_{-	1.09 	}^{+	1.55 	}$&$	2.32 	\pm	0.01 	$&$	63.69 	_{-	1.83 	}^{+	1.93 	}$&$	52.38 	_{-	0.93 	}^{+	0.92 	}$&	1070.94 	/	1275	&	1017	/	3829	/	4870	\\
501	&$	0.35 	\pm	0.02 	$&$	4.24 	_{-	1.36 	}^{+	2.27 	}$&$	2.44 	\pm	0.01 	$&$	73.04 	_{-	2.43 	}^{+	2.58 	}$&$	71.91 	_{-	1.27 	}^{+	1.26 	}$&	1167.57 	/	1275	&	1017	/	3979	/	5079	\\
601	&$	0.36 	\pm	0.02 	$&$	4.28 	_{-	1.46 	}^{+	2.58 	}$&$	2.41 	\pm	0.02 	$&$	68.30 	_{-	3.62 	}^{+	3.98 	}$&$	71.37 	_{-	1.71 	}^{+	1.70 	}$&	1113.24 	/	1275	&	479	/	3194	/	714	\\
901	&$	1.18 	\pm	0.01 	$&$ ^*	1.80 	\pm{	0.08 			}$&$	2.87 	\pm	0.01 	$&$^\dag			400.00 			$&$		144.77_{-	2.34 	}^{+	2.37 	}$&	1184.82 	/	1276	&	1676	/	2967	/	2467	\\
902	&$	1.17 	\pm	0.01 	$&$ ^*	1.91 	\pm{	0.08 			}$&$	2.87 	\pm	0.01 	$&$^\dag			400.00 			$&$		144.22_{-	2.33 	}^{+	2.37 	}$&	1210.98 	/	1276	&	1820	/	2780	/	3697	\\
903	&$	1.17 	\pm	0.01 	$&$ ^*	1.52 	\pm{	0.09 			}$&$	2.84 	\pm	0.01 	$&$^\dag			400.00 			$&$		143.93_{-	2.47 	}^{+	2.51 	}$&	1119.86 	/	1276	&	1116	/	2471	/	366	\\

		\hline
		\multicolumn{8}{|c|}{\bf{non-QPO}} \\
		144	&$	0.39 	\pm	0.01 	$&$	2.32 	_{-	0.48 	}^{+	0.66 	}$&$	2.24 	\pm	0.01 	$&$	67.18 	_{-	2.91 	}^{+	3.16 	}$&$	35.79 	_{-	0.84 	}^{+	0.85 	}$&	1166.41 	/	1275	&	913	/	1702	/	1250	\\
145	&$	0.38 	\pm	0.01 	$&$	2.48 	_{-	0.42 	}^{+	0.54 	}$&$	2.26 	\pm	0.01 	$&$	68.24 	_{-	1.89 	}^{+	1.99 	}$&$	37.72 	_{-	0.61 	}^{+	0.61 	}$&	1116.43 	/	1275	&	2531	/	6385	/	6977	\\
301	&$	0.36 	\pm	0.01 	$&$	4.30 	_{-	1.04 	}^{+	1.47 	}$&$	2.14 	\pm	0.01 	$&$	51.66 	_{-	1.79 	}^{+	1.91 	}$&$	35.61 	_{-	0.82 	}^{+	0.83 	}$&	1156.34 	/	1275	&	437	/	2400	/	1045	\\
401	&$	0.37 	\pm	0.01 	$&$	3.38 	_{-	0.78 	}^{+	1.09 	}$&$	2.30 	\pm	0.01 	$&$	61.91 	_{-	1.91 	}^{+	2.02 	}$&$	49.93 	_{-	0.99 	}^{+	1.00 	}$&	1067.75 	/	1275	&	792	/	2856	/	3521	\\
501	&$	0.37 	\pm	0.02 	$&$	2.92 	_{-	0.84 	}^{+	1.33 	}$&$	2.42 	\pm	0.01 	$&$	72.72 	_{-	2.60 	}^{+	2.77 	}$&$	69.07 	_{-	1.37 	}^{+	1.37 	}$&	1152.12 	/	1275	&	822	/	3015	/	3825	\\
601	&$	0.37 	\pm	0.02 	$&$	3.56 	_{-	1.18 	}^{+	2.06 	}$&$	2.40 	\pm	0.02 	$&$	69.12 	_{-	4.04 	}^{+	4.51 	}$&$	69.21 	_{-	1.85 	}^{+	1.85 	}$&	1132.31 	/	1275	&	396	/	2462	/	541	\\
901	&$	1.17 	\pm	0.01 	$&$ ^*	1.85 	\pm{	0.09 			}$&$	2.87 	\pm	0.01 	$&$^\dag			400.00 			$&$		143.77_{-	2.38 	}^{+	2.41 	}$&	1192.75 	/	1276	&	1605	/	2660	/	2234	\\
902	&$	1.17 	\pm	0.01 	$&$ ^*	1.96 	\pm{	0.09 			}$&$	2.87 	\pm	0.01 	$&$^\dag			400.00 			$&$		143.08_{-	2.36 	}^{+	2.41 	}$&	1228.03 	/	1276	&	1745	/	2496	/	3334	\\
903	&$	1.16 	\pm	0.01 	$&$ ^*	1.60 	\pm{	0.10 			}$&$	2.83 	\pm	0.01 	$&$^\dag			400.00 			$&$		142.53_{-	2.53 	}^{+	2.57 	}$&	1143.17 	/	1276	&	1068	/	2220	/	332	\\

		\hline
		\multicolumn{8}{|c|}{\bf{QPO}} \\
		144	&$	0.38 	\pm	0.03 	$&$	2.17 	_{-	0.76 	}^{+	1.32 	}$&$	2.30 	\pm	0.02 	$&$	72.17 	_{-	5.02 	}^{+	5.69 	}$&$	41.52 	_{-	1.53 	}^{+	1.53 	}$&	1180.61 	/	1213	&	220	/	456	/	425	\\
145	&$	0.35 	\pm	0.02 	$&$	3.85 	_{-	1.08 	}^{+	1.65 	}$&$	2.29 	\pm	0.01 	$&$	69.05 	_{-	2.55 	}^{+	2.73 	}$&$	42.03 	_{-	0.91 	}^{+	0.91 	}$&	1043.74 	/	1275	&	635	/	1777	/	2304	\\
301	&$	0.33 	\pm	0.02 	$&$	8.88 	_{-	3.52 	}^{+	6.57 	}$&$	2.18 	\pm	0.02 	$&$	54.27 	_{-	2.38 	}^{+	2.56 	}$&$	40.01 	_{-	1.15 	}^{+	1.16 	}$&	1110.16 	/	1204	&	159	/	965	/	582	\\
401	&$	0.33 	\pm	0.02 	$&$	7.44 	_{-	3.10 	}^{+	6.10 	}$&$	2.34 	\pm	0.02 	$&$	66.63 	_{-	2.83 	}^{+	3.04 	}$&$	56.65 	_{-	1.45 	}^{+	1.44 	}$&	1100.27 	/	1239	&	220	/	951	/	1322	\\
501	&$	0.32 	\pm	0.04 	$&$	8.22 	_{-	4.90 	}^{+	16.97 	}$&$	2.43 	\pm	0.02 	$&$	71.33 	_{-	3.61 	}^{+	3.93 	}$&$	76.11 	_{-	2.11 	}^{+	2.06 	}$&	1116.32 	/	1230	&	190	/	939	/	1221	\\
601	&$	0.34 	\pm	0.05 	$&$	4.56 	_{-	2.81 	}^{+	10.92 	}$&$	2.43 	\pm	0.03 	$&$	70.41 	_{-	6.53 	}^{+	7.70 	}$&$	76.53 	_{-	3.09 	}^{+	3.03 	}$&	1086.69 	/	1129	&	80	/	712	/	169	\\
901	&$	1.22 	\pm	0.02 	$&$ ^*	1.53 	\pm{	0.20 			}$&$	2.84 	\pm	0.02 	$&$^\dag			400.00 			$&$		145.38_{-	6.27 	}^{+	6.36 	}$&	1037.49 	/	1093	&	63	/	278	/	210	\\
902	&$	1.18 	\pm	0.02 	$&$ ^*	1.79 	\pm{	0.23 			}$&$	2.84 	\pm	0.02 	$&$^\dag			400.00 			$&$		144.24_{-	6.17 	}^{+	6.33 	}$&	1212.58 	/	1104	&	67	/	257	/	327	\\
903	&$	1.23 	\pm	0.03 	$&$ ^*	1.21 	\pm{	0.23 			}$&$	2.81 	\pm	0.02 	$&$^\dag			400.00 			$&$		145.46_{-	7.20 	}^{+	7.27 	}$&	1071.39 	/	1050	&	43	/	228	/	31	\\

		\hline
		\multicolumn{8}{|l|}{$^*$: ($10^3$)} \\
		\multicolumn{8}{|l|}{$^\dag$: fixed} \\
	\end{tabular}
%	\begin{tablenotes}
%		\footnotesize
%		\item[1] time-averaged
%	\end{tablenotes}
%	\end{threeparttable}
\end{table*}

\section{Spectral Results}

The reported QPOs in Table~\ref{tab:QPO} are all type-C QPOs \citep{Huang2018}, and Obs 144, 145, 301, 401, 501 and 601 are in HIMS, while 901, 902 and 903 are in SIMS. As seen in Table~\ref{tab:xspec}, the inner disk temperature in SIMS is hotter than that in HIMS, raised by around 1.0 keV, and the change is abrupt. The photon index $\Gamma$, however, shows relatively gradual evolution from HIMS to SIMS, increased by about 0.5. Our main aim here is to study the different behaviors of QPO and non-QPO spectra. To describe the differences between the QPO and non-QPO spectra in a well defined way, the fitting parameters of the QPO/non-QPO spectra were normalized to the corresponding fitting parameters of the time-averaged spectra. In this work, the ratio of the fitting parameters for the QPO/non-QPO spectra to the fitting parameters for the time-averaged spectra, are labeled with a letter $\rm r$. For example, $\Gamma^{\rm r}$ corresponds to the ratio of the QPO/non-QPO spectrum index to the time-averaged spectrum index.

\begin{figure}
	\includegraphics[width=\columnwidth]{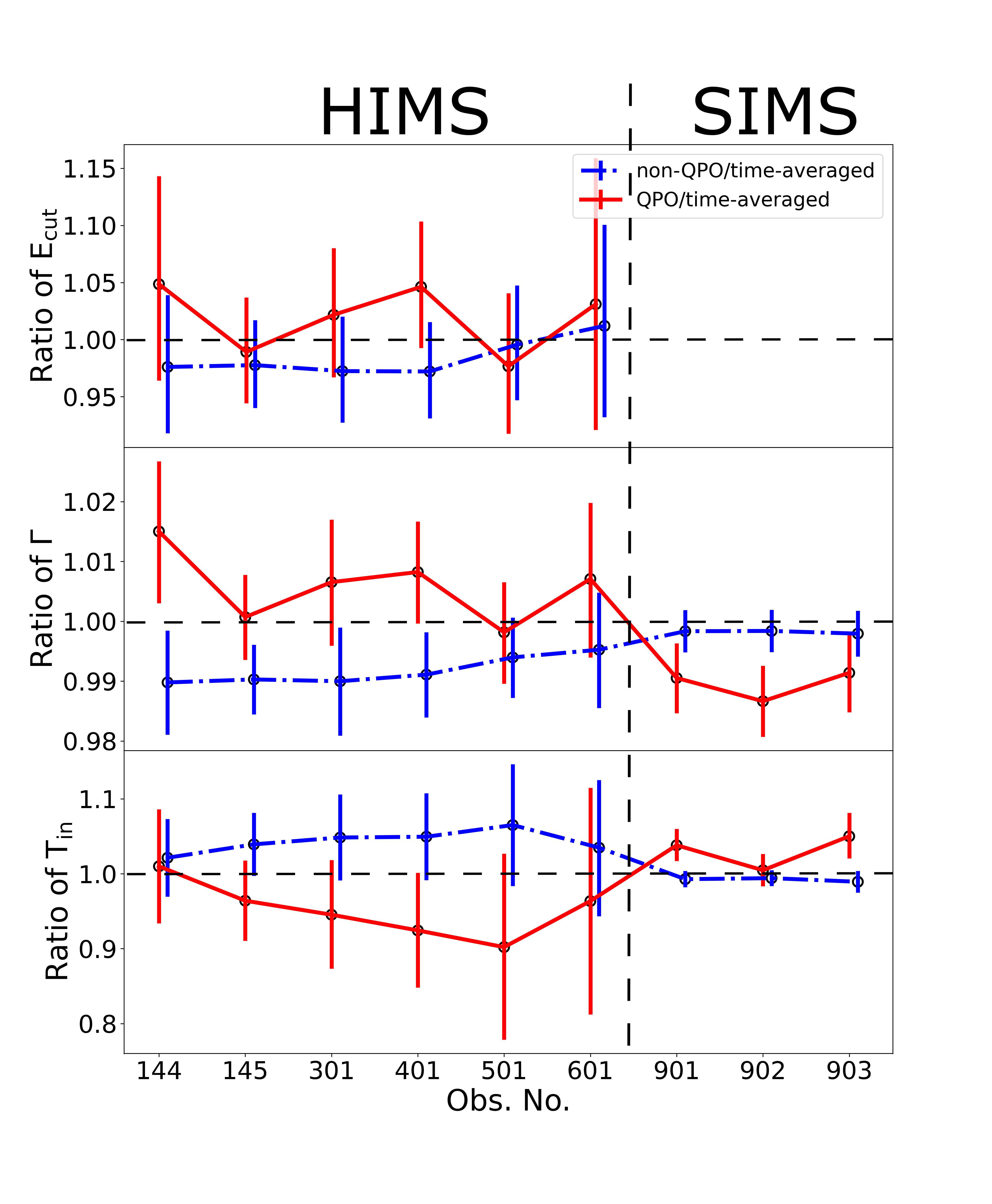}
	\caption{The spectral fitting parameters of nine observations. The model used is const*tbabs*(diskbb+cutoffpl), with $N_H$ fixed to $3.8\times 10^{22} \rm \ cm^{-2}$. The vertical axes are the ratio of the QPO/non-QPO fitting parameters to the corresponding time-averaged values. From top to bottom, cut-off energy $E_{\rm cut}^r$, photon index $\Gamma^r$, and inner disc temperature $T_{\rm in}^r$ are presented respectively. The solid red lines present the QPO regime, while the blue dash-dot lines show the non-QPO regime. The non-QPO points are shifted a bit to the right to clearly show the ranges of error bars.}
	\label{fig:parameters}
\end{figure}

In Figure~\ref{fig:parameters}, we show the best-fitting spectral parameters of the QPO and non-QPO spectra including the photon index $\Gamma^r$, inner disk temperature $T_{\rm in}^r$ and the cutoff energy $E_{\rm cut}^r$, which indicate the evolution with the spectral transition. The QPO spectra are softer than the non-QPO spectra in HIMS, while in SIMS, the QPO spectra are slightly harder. For the inner disk temperature, although the error bars overlap for most observations, the difference may still exist: the disk temperature is lower for the QPO case in HIMS, but becomes similar with the non-QPO case in SIMS. There is no difference for the cutoff energy between QPO and non-QPO spectra.

%In HIMS, the fitting parameters of QPO spectra show a lower inner disk temperature ($\sim 10 \%$) and a higher photon index ($\sim 2 \%$) than the non-QPO ones, with a clear cutoff energy. While in SIMS, higher temperature ($\sim 3 \%$) and lower photon index ($< 1 \%$) of QPO spectra are presented with unconstrained cutoff energy.

%For disk component, most of ME and HE fitting cannot be well constrained, and the remaining few fittings give values around $-11 cgs$ in logarithm, thus only LE results are taken as the final disk flux.

\begin{figure}
	\includegraphics[width=\columnwidth]{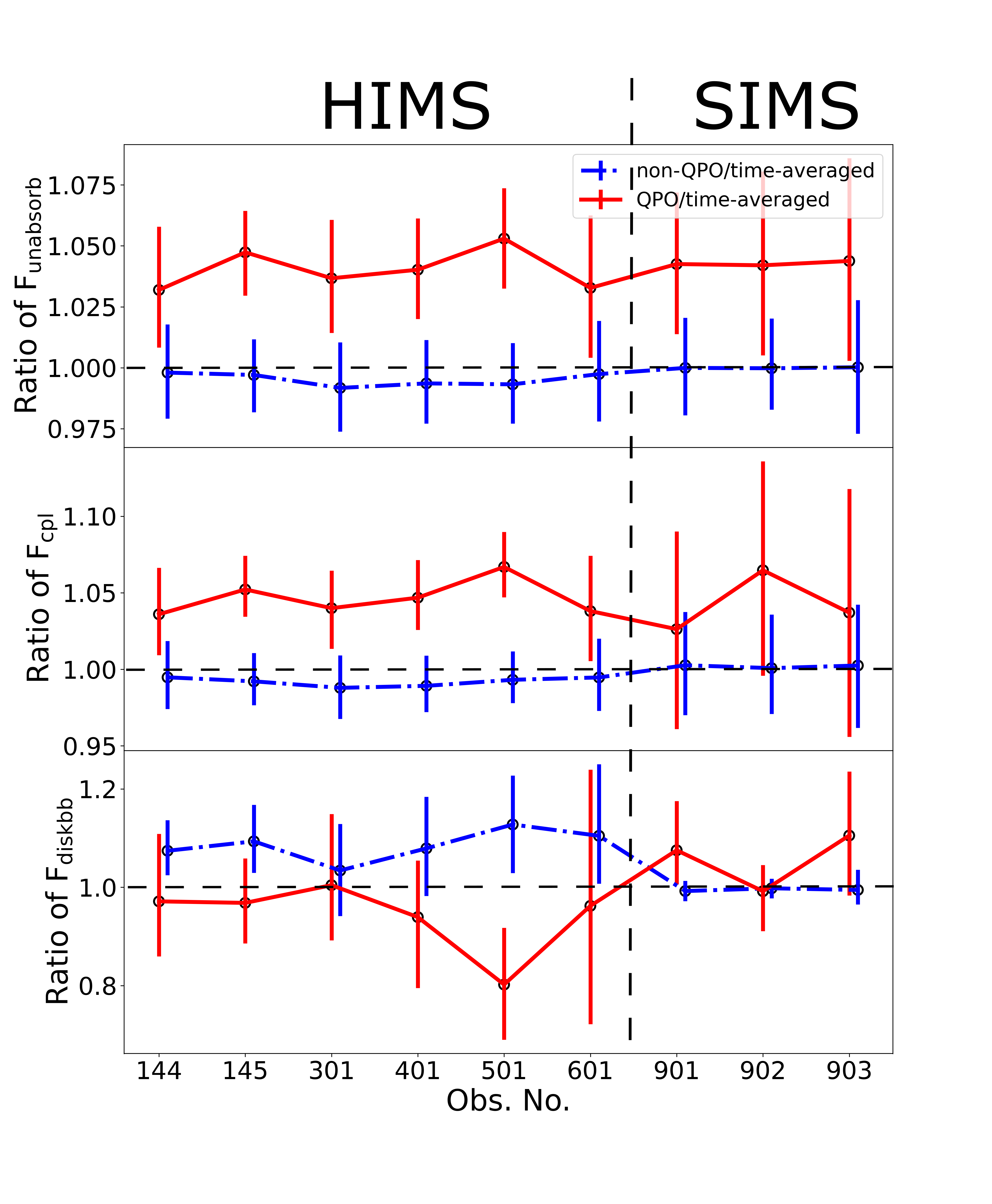}
	\caption{The summed flux of 2-100 keV with error bars. The vertical axes are the ratio of the QPO/non-QPO flux to the corresponding time-averaged flux. From top to bottom, total flux $F_{\rm unabsorb}^r$, cutoff powerlaw flux $F_{\rm cpl}^r$, and disk flux $F_{\rm disk}^r$ are shown respectively. The solid red lines present the QPO regime, while the blue dash-dot lines show the non-QPO regime. The non-QPO points are shifted a bit to the right to clearly show the ranges of error bars.}
	\label{fig:3flux}
\end{figure}

In order to understand the flux evolution of the spectral components, we have estimated the flux of each model component by using cflux in XSPEC. We separately fit LE (2-10 keV), ME (10-27 keV) and HE (27-100 keV) spectra for each component, and add them together. Table~\ref{tab:flux} presents the summed flux of 2-100 keV, and Figure~\ref{fig:3flux} shows the ratio of the QPO/non-QPO flux to the time-averaged flux. Both QPO and non-QPO unabsorbed fluxes $F_{\rm unabsorb}^r$ show a similar trend, except the flux of QPO regime is a little higher than the non-QPO flux although some values have large errors. The powerlaw flux $F_{\rm cpl}^r$ exhibits an approximate trend to $F_{\rm unabsorb}^r$, except in SIMS, the overlap between QPO and non-QPO regimes is more obvious. As for the disk flux $F_{\rm disk}^r$, it is relatively steady throughout the nine observations during the non-QPO regime, but becomes quite different in the QPO regime. $F_{\rm disk}^r$ of QPO spectra is lower than that of non-QPO spectra in HIMS, but has a sudden rising from HIMS to SIMS (the flux $F_{\rm disk}$ also rises $\sim$ one order of magnitude from Obs 601 to 901 as seen in Table~\ref{tab:flux}), and becomes approximately or higher than the non-QPO $F_{\rm disk}^r$ when the source transits to SIMS. % Meanwhile, $F_{\rm cpl}^r$ of the QPO spectra is higher than that of the non-QPO spectra in HIMS, but becomes lower in SIMS.

\begin{table}
	\centering
	\caption{The 2-100 keV summed flux values of the diskbb component, the cutoffpl component, and the total unabsorbed component.}
	\label{tab:flux}
	\begin{tabular}{cccc} % four columns, alignment for each
		\hline
		Obs & $F_{disk} (10^{-9})$ & $F_{cpl} (10^{-9})$ & $F_{unabsorb} (10^{-9})$\\
		& ($erg/cm^2/s$)& ($erg/cm^2/s$)& ($erg/cm^2/s$) \\
		\hline
		\multicolumn{4}{|c|}{\bf{time-averaged}} \\
144 & $	4.95 	_{-	0.18 	}^{+	0.23 	}$&$	97.47 	_{-	1.34 	}^{+	1.58 	}$&$	102.81 	_{-	1.29 	}^{+	1.38 	}$	\\
145 & $	4.90 	_{-	0.20 	}^{+	0.23 	}$&$	98.77 	_{-	1.03 	}^{+	1.36 	}$&$	104.03 	_{-	1.16 	}^{+	1.03 	}$	\\
301 & $	6.21 	_{-	0.37 	}^{+	0.38 	}$&$	113.68 	_{-	1.59 	}^{+	1.69 	}$&$	119.99 	_{-	1.39 	}^{+	1.72 	}$	\\
401 & $	5.68 	_{-	0.37 	}^{+	0.38 	}$&$	120.50 	_{-	1.39 	}^{+	1.69 	}$&$	126.51 	_{-	1.59 	}^{+	1.58 	}$	\\
501 & $	4.39 	_{-	0.27 	}^{+	0.28 	}$&$	133.78 	_{-	1.00 	}^{+	1.76 	}$&$	139.11 	_{-	1.55 	}^{+	1.75 	}$	\\
601 & $	5.00 	_{-	0.25 	}^{+	0.57 	}$&$	138.09 	_{-	2.07 	}^{+	2.41 	}$&$	143.60 	_{-	1.92 	}^{+	2.25 	}$	\\
901 & $	41.50 	_{-	0.61 	}^{+	0.61 	}$&$	135.75 	_{-	3.13 	}^{+	3.11 	}$&$	177.64 	_{-	2.29 	}^{+	2.43 	}$	\\
902 & $	43.09 	_{-	0.60 	}^{+	0.59 	}$&$	134.58 	_{-	2.84 	}^{+	2.96 	}$&$	177.70 	_{-	2.20 	}^{+	2.43 	}$	\\
903 & $	33.39 	_{-	0.73 	}^{+	1.08 	}$&$	142.94 	_{-	3.95 	}^{+	3.95 	}$&$	177.91 	_{-	3.36 	}^{+	3.39 	}$	\\

		\hline
		\multicolumn{4}{|c|}{\bf{non-QPO}} \\
144 & $	5.32 	_{-	0.15 	}^{+	0.19 	}$&$	96.96 	_{-	1.51 	}^{+	1.70 	}$&$	102.61 	_{-	1.45 	}^{+	1.50 	}$	\\
145 & $	5.36 	_{-	0.23 	}^{+	0.26 	}$&$	97.99 	_{-	1.14 	}^{+	1.23 	}$&$	103.73 	_{-	1.09 	}^{+	1.13 	}$	\\
301 & $	6.42 	_{-	0.43 	}^{+	0.44 	}$&$	112.31 	_{-	1.69 	}^{+	1.74 	}$&$	118.99 	_{-	1.65 	}^{+	1.45 	}$	\\
401 & $	6.12 	_{-	0.38 	}^{+	0.43 	}$&$	119.20 	_{-	1.54 	}^{+	1.70 	}$&$	125.69 	_{-	1.36 	}^{+	1.63 	}$	\\
501 & $	4.96 	_{-	0.31 	}^{+	0.30 	}$&$	132.86 	_{-	1.78 	}^{+	1.78 	}$&$	138.16 	_{-	1.63 	}^{+	1.60 	}$	\\
601 & $	5.53 	_{-	0.40 	}^{+	0.37 	}$&$	137.36 	_{-	2.21 	}^{+	2.57 	}$&$	143.22 	_{-	2.03 	}^{+	2.19 	}$	\\
901 & $	41.18 	_{-	0.60 	}^{+	0.61 	}$&$	136.11 	_{-	3.12 	}^{+	3.54 	}$&$	177.63 	_{-	2.59 	}^{+	2.72 	}$	\\
902 & $	43.01 	_{-	0.63 	}^{+	0.59 	}$&$	134.68 	_{-	2.85 	}^{+	3.67 	}$&$	177.67 	_{-	2.06 	}^{+	2.70 	}$	\\
903 & $	33.21 	_{-	0.66 	}^{+	0.85 	}$&$	143.30 	_{-	4.25 	}^{+	4.07 	}$&$	177.94 	_{-	3.49 	}^{+	3.53 	}$	\\

		\hline
		\multicolumn{4}{|c|}{\bf{QPO}} \\
144 & $	4.81 	_{-	0.53 	}^{+	0.64 	}$&$	100.97 	_{-	2.20 	}^{+	2.47 	}$&$	106.09 	_{-	2.03 	}^{+	2.25 	}$	\\
145 & $	4.74 	_{-	0.36 	}^{+	0.38 	}$&$	103.92 	_{-	1.37 	}^{+	1.64 	}$&$	108.96 	_{-	1.39 	}^{+	1.41 	}$	\\
301 & $	6.23 	_{-	0.59 	}^{+	0.81 	}$&$	118.22 	_{-	2.51 	}^{+	2.17 	}$&$	124.39 	_{-	2.27 	}^{+	2.26 	}$	\\
401 & $	5.33 	_{-	0.74 	}^{+	0.55 	}$&$	126.14 	_{-	2.06 	}^{+	2.37 	}$&$	131.59 	_{-	1.95 	}^{+	2.09 	}$	\\
501 & $	3.52 	_{-	0.44 	}^{+	0.45 	}$&$	142.72 	_{-	2.43 	}^{+	2.43 	}$&$	146.48 	_{-	2.34 	}^{+	2.21 	}$	\\
601 & $	4.81 	_{-	1.18 	}^{+	1.28 	}$&$	143.34 	_{-	3.95 	}^{+	4.31 	}$&$	148.30 	_{-	3.60 	}^{+	3.59 	}$	\\
901 & $	44.63 	_{-	2.80 	}^{+	4.10 	}$&$	139.31 	_{-	8.26 	}^{+	8.06 	}$&$	185.20 	_{-	4.50 	}^{+	4.57 	}$	\\
902 & $	42.75 	_{-	3.44 	}^{+	2.22 	}$&$	143.29 	_{-	8.76 	}^{+	9.05 	}$&$	185.17 	_{-	6.15 	}^{+	6.35 	}$	\\
903 & $	36.91 	_{-	4.00 	}^{+	4.18 	}$&$	148.24 	_{-	10.84 	}^{+	10.79 	}$&$	185.70 	_{-	6.37 	}^{+	6.61 	}$	\\

		\hline
	\end{tabular}
\end{table}

\begin{figure}
	\includegraphics[width=\columnwidth]{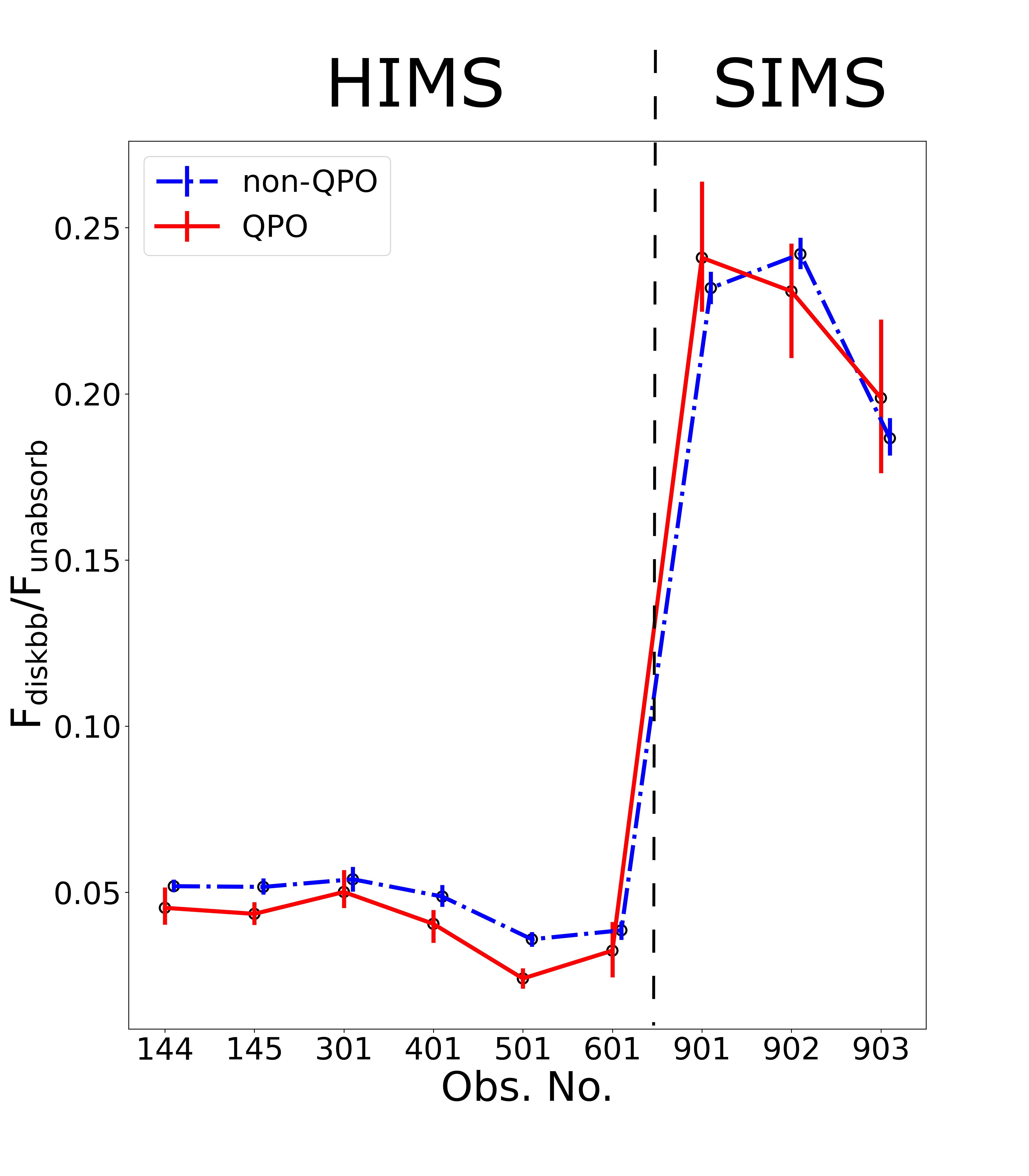}
	\caption{The proportion of the disk flux component defined by the ratio of $F_{\rm disk}$ to $F_{\rm unabsorb}$. The solid red line is the QPO disk proportion, while the blue dash-dot line is the non-QPO regime. The non-QPO points are shifted a bit to the right to clearly show the ranges of error bars.}
	\label{fig:flux}
\end{figure}

The ratio of $F_{\rm disk}$ to $F_{\rm unabsorb}$ is plotted in Figure~\ref{fig:flux} to show the evolution of the disk flux. Both QPO and non-QPO regimes show a sudden increase during the evolution from HIMS to SIMS. For the QPO regime, this ratio is lower in HIMS compared to non-QPO regime, but increases faster between Obs 601-901, and becomes similar to the ratio of the non-QPO regime in SIMS.

\begin{figure}
	\includegraphics[width=\columnwidth]{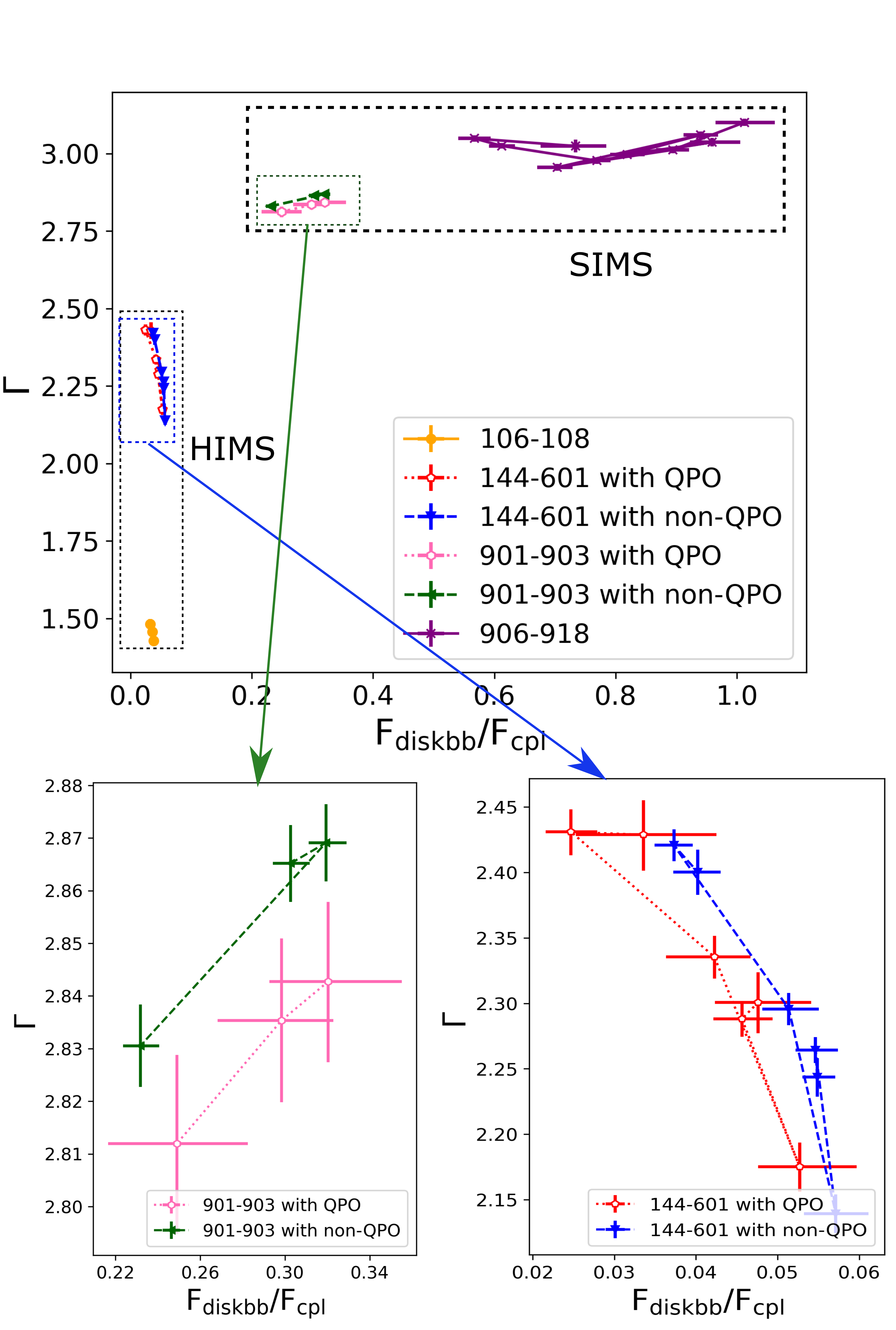}
	\caption{Photon index versus the ratio of the disk flux to the cutoff power-law flux ($top panel$). Both QPO and non-QPO points are presented for Obs 144--601 (red pentagon and blue inverted triangle, respectively) and Obs 901--903 (pink hexagon and green left-facing triangle, respectively). The time-averaged results of Obs 106--108 (orange circle) and 906--918 (purple cross) without QPO signal detected are also plotted. The diagrams for Obs 144--601 in HIMS and Obs 901--903 in SIMS are also zoomed in for the better presentation of the QPO and non-QPO cases ($bottom panel$).}
	\label{fig:fluxGammaRatio}
\end{figure}

Moreover, we plot $\Gamma$ and $F_{\rm disk} / F_{\rm cpl}$ in Figure~\ref{fig:fluxGammaRatio} for Obs 106-918, where 106-108 and 906-918 are time-averaged data since they have no QPO signals detected, and 144-903 include the QPO and non-QPO separated data. QPO data and non-QPO data have shown diverse relations in different states. During HIMS (see the bottom right panel of Figure~\ref{fig:fluxGammaRatio}), a Spearman Rank Correlation test shows that, the Spearman rank correlation $r$ is -0.918 and the corresponding p-value $p= 0.009$ for the QPO regime, and $r=-0.940$ with $p=0.005$ for the non-QPO regime, indicating that the photon index has a negative relation to the ratio of $F_{\rm disk}/F_{\rm cpl}$. In addition, the QPO data have smaller $F_{\rm disk}/F_{\rm cpl}$ and slightly softer spectra compared to the non-QPO data. However, the different relation appears in the case of SIMS (see the bottom left panel of Figure~\ref{fig:fluxGammaRatio}), with $r=0.863$ and $p=0.33$, $r=0.970$ and $p=0.15$ for the QPO and non-QPO regimes respectively, and the QPO spectra are harder and the disk components are becoming similar to the non-QPO data. Because there are only three data points in SIMS, this marginal relation between $\Gamma$ and the flux ratio needs more observations to confirm.

\section{Discussions \& Conclusions}

In this work, we have applied the wavelet method to the QPO analysis in X-ray binaries, and the time-frequency evolution of the type-C QPOs in MAXI J1535-571 has been studied. The type-C QPOs show the transient signals even in the short time scales (down to several seconds). With the wavelet method, we can separate the QPO and non-QPO events and derive the spectra for the QPO and non-QPO regimes, respectively. The wavelet results revealed the transitions between the QPO and non-QPO signals throughout the whole observations from the HIMS to SIMS. The different spectral properties between QPO and non-QPO regimes may help to probe the origin of the QPO near BH.

\cite{Xu2019} reported a sudden appearance of type-C QPO along with its flux decreasing by $\sim 45 \%$ in Swift J1658.2-4242. After comparing accumulated spectra before and after the flux change, they found that both disk and corona component emission decreased with its disk temperature dropping by $\sim 15 \%$, but the inner disk radius and the coronal properties almost remained invariable. In the analysis of MAXI J1535-571 when QPO is present, the diskbb norm parameter also shows no notable discrepancy in the SIMS, indicating the fact that the inner edge of the accretion disk is stable (see Table~\ref{tab:xspec}). The disk temperature $T_{\rm in}$ of the QPO regime during the HIMS may systematically decrease ($\sim 10 \%$ in average though large error bars exist, see Table~\ref{tab:xspec}) compared with the non-QPO regime, but there may also be a possibility that the non-QPO time intervals in the wavelet analysis may contain some possible QPO time intervals with the very low statistics for the QPO signal (e.g., below $95 \%$ confidence level), $T_{\rm in}$ would not change significantly for two regimes defined here. Moreover, instead of remaining unchanged in Swift J1658.2-4242, the coronal property $\Gamma^{\rm r}$ in our source shows a relatively more pronounced variation as compared to $T_{\rm in}^{\rm r}$ (see Figure~\ref{fig:parameters}). Our total flux values also do not show the same trend as in the source Swift J1658.2-4242. On the contrary, the total flux in the QPO regimes was higher as shown in Figure~\ref{fig:3flux}. The sudden QPO appearance in Swift J1658.2-4242 may not have the similar origin with the fast evolution of the QPO phenomenon in MAXI J1535-571.

\cite{Zhang2021} analyzed series of fast appearances and disappearances of type-B QPOs in SIMS in MAXI J1348-30. They found that the spectral difference between the QPO and non-QPO regimes became significant above 2 keV, although subtle difference existed below 2 keV (see their Figure 6). As a comparison, our 2-100 keV spectra in the SIMS (e.g. the bottom panel of Figure~\ref{fig:example_3combo}) show a similar discrepancy. The difference between the QPO and non-QPO spectra also gradually became apparent, and the higher the energy, the greater the difference. However during HIMS, the spectral difference of our data still existed in LE, but no obvious discrepancy can be noted in the ME and HE (see the top panel of Figure~\ref{fig:example_3combo}). Because the energy range of \cite{Zhang2021} is only up to 10 keV, we are unable to assert whether the part above 10 keV is similar to our SIMS spectra, or is consistent with our HIMS spectra. For the flux comparison, their disk and Comptonized components varied with the same behavior as our HIMS regime. Though type-B and type-C QPOs would have different physical origins, the transition between QPO and non-QPO regimes may share similar physics.

The spectral results on QPO and non-QPO regimes in this wavelet work combined with the QPO type work suggested that the differences between HIMS and SIMS may be due to the sudden appearance of type-B QPO in MAXI J1535-571. The type-C QPOs were studied in both HIMS and SIMS of MAXI J1535-571, and there are differences between QPO and non-QPO spectral behaviours from HIMS to SIMS. The Insight-HXMT also found strong type-B QPO signal just in the first $\sim$ 800 s of dynamical PDS in Obs 701 \citep{Huang2018}, which just occurred around the dates between HIMS and SIMS in our analysis. \cite{Sriram2021} reported a type-C to type-B transformation event in the source H1743-322, and the two-component hot flow model fitting suggested the fundamental QPO and the harmonic were excited by an inner hot flow/jet and the outer part of a hot Compton component, respectively. They speculated that during the transition from type-C to B, the outer hot flow was ejected away and only the inner part was left. This conjecture may explain our results on the spectral variation from HIMS to SIMS. There seems to be a visible increase in both the disk temperature and flux during the evolution from HIMS to SIMS in our data (see Tables~\ref{tab:xspec} and \ref{tab:flux}). If somehow the outer flow ejection happened and covered part of the disk, then the departure of the outer part would cause the reduction of the portion of the disk photons being Comptonized by the hot flow, leading to the increase of both the disk flux and temperature.

It was thought that the observed QPO may originate from the Lense-Thirring precession of the inner hot flow \citep{Ingram2009,You2018,You2020}. In this scenario, the ratio of the disk luminosity intercepted by the inner hot flow to the total disk emitted luminosity becomes maximum, when the inner hot flow aligns with the disk. 
Those intercepted photons will be inverse-Compton scattered in the precessing inner flow and illuminate back to the outer disk. Therefore, given the total disk emitted luminosity, we would expect more Comptonized disk photons to illuminate and heat up the disk, when the flow aligns with the outer disk (see the upper panel of Figure 2 in \citealt{You2018}, also \citealt{Zycki2016,Ingram2019}), which would lead to the increase of the disk temperature.
%In other words, the disk temperature will be higher in this case, providing that the disk is heated by the illumination of the inner hot flow and the viscous dissipation.
Therefore, for the non-QPO time intervals during HIMS when the precession is highly suppressed, the inner hot flow aligns with the disk, which would make the disk hotter. Such a scenario of the precession could also explain the harder non-QPO spectra in HIMS. For the time intervals with non-QPO, i.e. the inner hot flow being aligning with the outer disk, the disk luminosity is highly intercepted by the inner hot flow. This will then lead to the increase of the Comptonization flux from the hot flow and the decrease of the disk flux. However, the hotter disk will create more radiation, which causes the total disk flux to increase or remain almost unchanged. The increased Comptonization flux may then be affected by the geometric change of the corona, or by a partial ejection of the Comptonized region as described in \cite{Sriram2021} and \cite{Zhang2021}, resulting in a decrease in the line-of-sight flux, i.e. $F_{\rm cpl}$.

In summary, we have studied the light curves of MAXI J1535-571 using wavelet analysis with the Insight-HXMT data. Nine observations are detected with QPO signals, and their spectra are divided into QPO and non-QPO regime based on wavelet results. The QPO spectra and non-QPO spectra show different relations in evolution, which means their physical mechanism may not be the same. In HIMS, the QPO regime has lower disk temperature, softer spectra and lower disk flux compared to the non-QPO regime. These relations are reversed in SIMS. The Comptonization flux and total flux in QPO spectra however, are always higher regardless of HIMS or SIMS.

The reversed relation between HIMS and SIMS may be related to the transient appearance of type-B QPO in MAXI J1535-571 based on the wavelet results. A type-C to type-B transition was also reported by \cite{Sriram2021}, but to analyze the connection and process between the two intermediate states, a complete C-B-C process analysis is needed with more observational data. Even though we cannot draw very convincing conclusions on the origin of QPO and non-QPO spectral behaviours due to insufficient data, our results still reveal that wavelet analysis utilized in QPO study is an effectively and promising method, and can be attempted for future QPO research.

\section*{Acknowledgements}

We are very grateful to the referee for the fruitful suggestions to improve this manuscript. This work is supported by the National Key Research and Development Program of China (Grants No. 2021YFA0718500, 2021YFA0718503), the NSFC (12133007, U1838103, 11622326, U1838201, U1838202, 11903024, U1931203), and the Fundamental Research Funds for the Central Universities (No. 2042021kf0224). This work made use of data from the \textit{Insight}-HXMT mission, a project funded by China National Space Administration (CNSA) and the Chinese Academy of Sciences (CAS).

%%%%%%%%%%%%%%%%%%%%%%%%%%%%%%%%%%%%%%%%%%%%%%%%%%
\section*{Data Availability}
Data that were used in this paper are from Institute of High Energy Physics Chinese Academy of Sciences(IHEP-CAS) and are publicly available for download from the \textit{Insight}-HXMT website.
To process and fit the spectrum and obtain folded light curves, this research has made use of XRONOS and FTOOLS provided by NASA.



\bibliographystyle{mnras}
\bibliography{ref} % if your bibtex file is called example.bib

%%%%%%%%%%%%%%%%%%%%%%%%%%%%%%%%%%%%%%%%%%%%%%%%%%

%%%%%%%%%%%%%%%%% APPENDICES %%%%%%%%%%%%%%%%%%%%%

\appendix

\section{Comparison of dynamical power spectrum and wavelet analysis}
To extract time-frequency information from a signal, two possible methods can be performed, one is the windowed Fourier transform (WFT) and the other is the wavelet transform. WFT is often used in the research of QPO signals, and the dynamical PDS generated by this technique is very helpful to analyze the sudden occurrence and disappearance of QPOs. However, this method has its drawbacks. First of all, the WFT is inefficient since the $T/2\delta t$ frequencies must be analyzed at each time step, where $T$ is the length of a sliding segments \citep{Torrence1998}. Secondly, as indicated by \cite{Kaiser2011}, the dividing line between time and frequency is determined by the choice of $T$, thus we cannot assert whether the selected window contains exactly a certain frequency component. As a consequence, several $T$ must be tested, but it is still possible to miss the appropriate results, which is tough to make new discoveries. Wavelet analysis, as the other method, can solve both of the above problems because this method is scale independent. In the meanwhile, the global wavelet spectrum provided by the wavelet method is time-averaged, so it is smooth, unbiased and consistent, rather than full of noise and false peaks like the Fourier spectrum.

In order for readers to compare, we show an example of the dynamical PDS results below. The data utilized in this plot are exactly the same as Figure~\ref{fig:example_wa_amplify}, with time resolution fixed at 1 s. RMS normalization \citep{Miyamoto1991} is applied on the power. Compared with the wavelet plot, the WFT technique shows similar peak positions and numbers, except the time-frequency resolution is much lower.

\begin{figure}
	\includegraphics[width=\columnwidth]{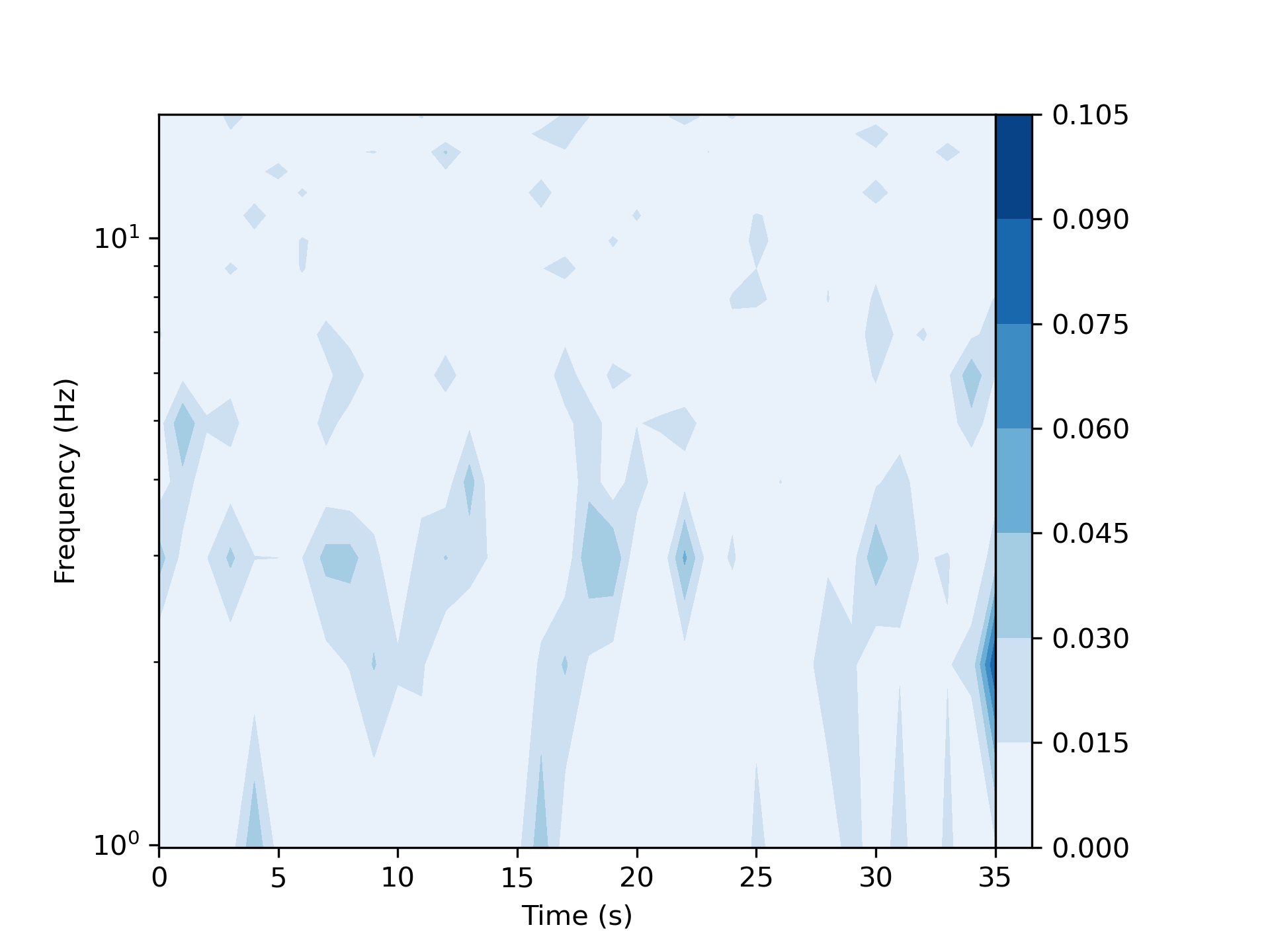}
	\caption{The dynamic PDS results derived with the same data in Figure~\ref{fig:example_wa_amplify}. RMS normalization is applied on the power. Darker color means higher power.}
	\label{fig:DynamicalPDS}
\end{figure}

\begin{figure}
	\begin{minipage}{0.5\textwidth}
		\includegraphics[width=1\textwidth]{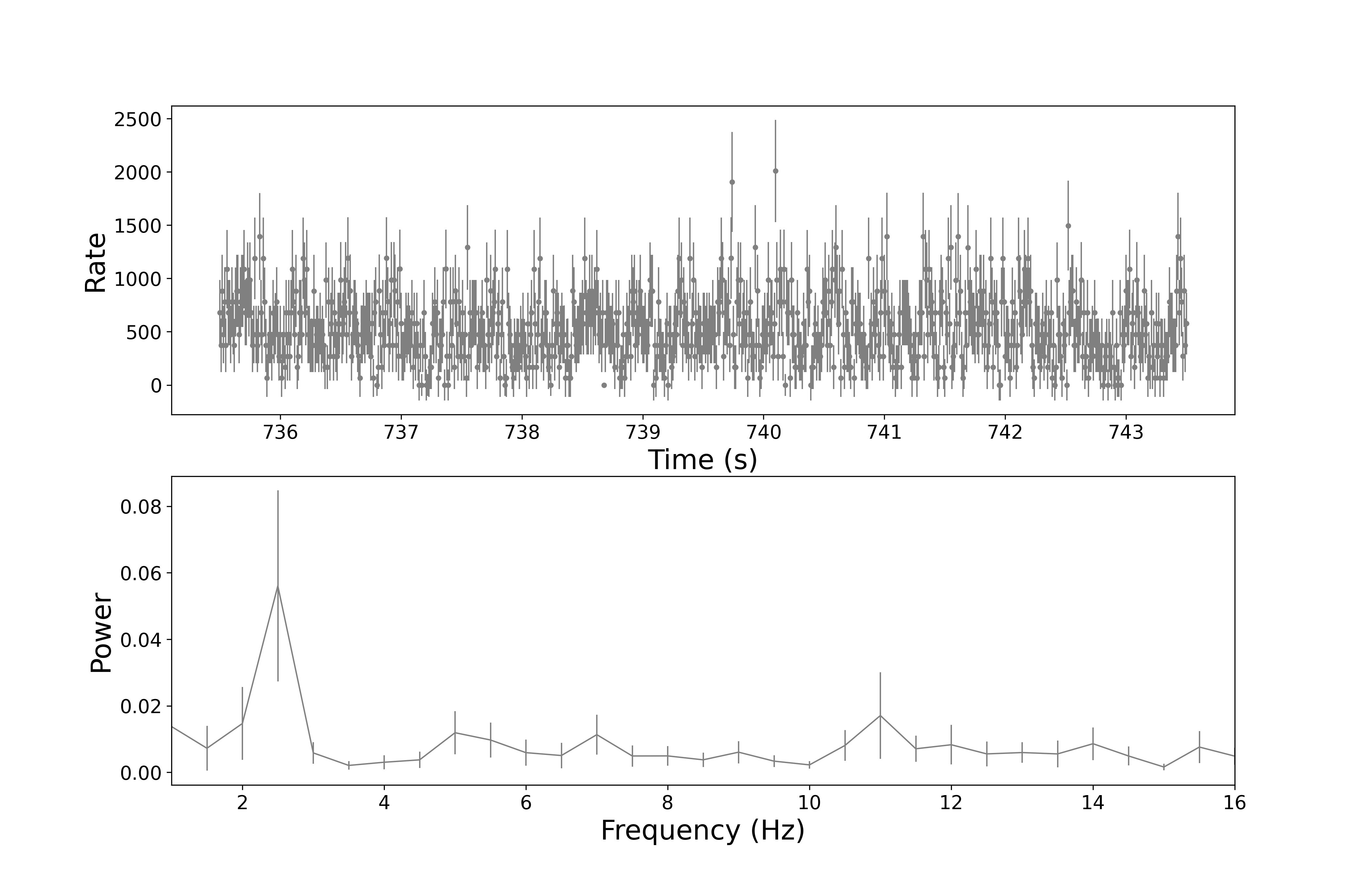}
	\end{minipage}
	\begin{minipage}{0.5\textwidth}
		\includegraphics[width=1\textwidth]{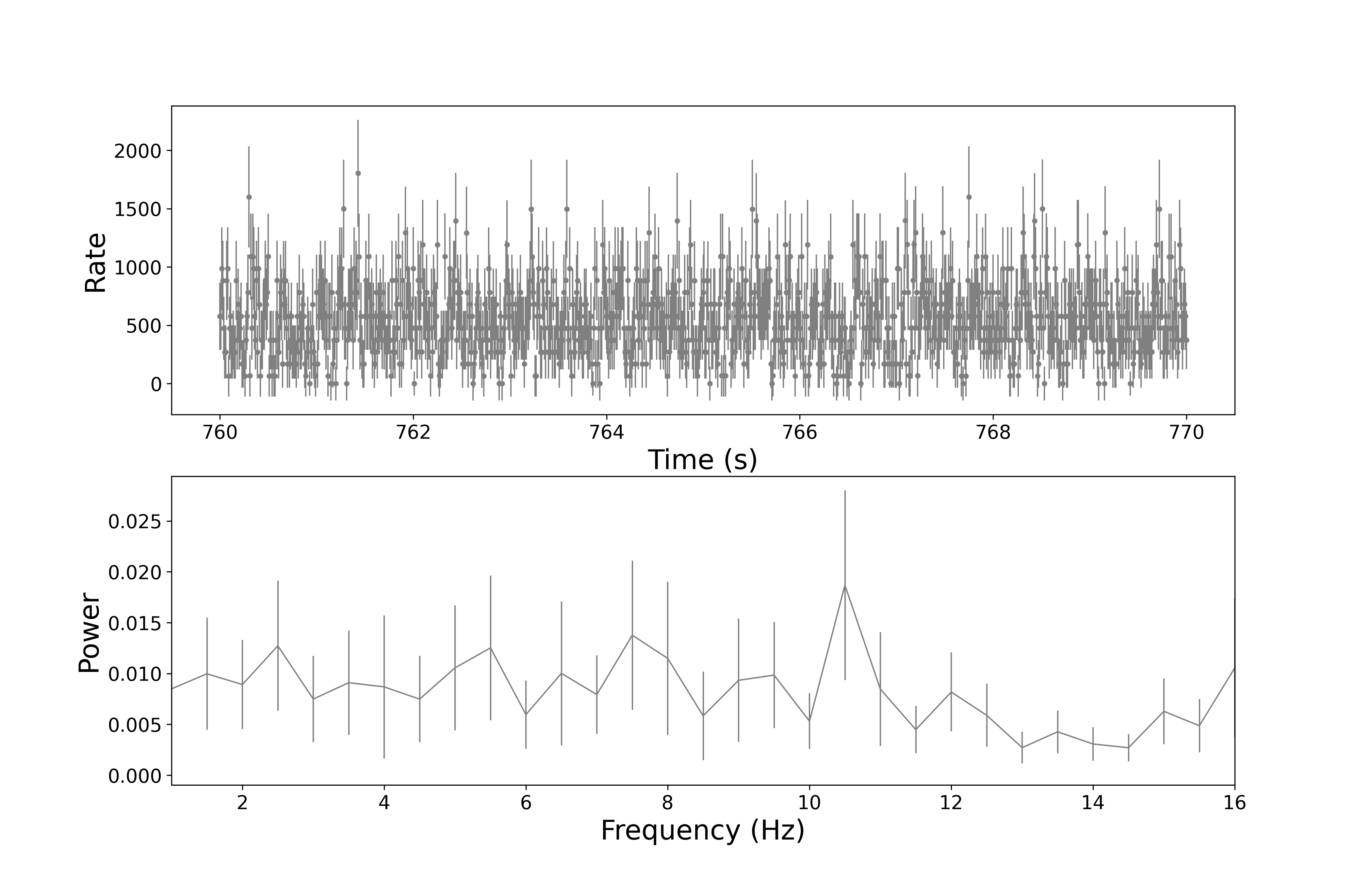}
	\end{minipage}
	\caption{The examples of the lightcurves and corresponding power spectra for the time intervals of QPO (top panel) and non-QPO (bottom panel) identified by our wavelet method.}
	\label{fig:fftNlc}
\end{figure}

To further confirm our wavelet result, we checked the light curve data for QPO and non-QPO time resolved by wavelet analysis, and Fourier transforms for the different light curves are performed for comparison. In Figure~\ref{fig:fftNlc}, the QPO light curve and PDS are presented on the top panel, while non-QPO light curve and corresponding PDS on the bottom panel. The PDS and the error of power are produced from 2-second data intervals with RMS normalization applied. The error of power relies on the number of power averaged in each bin, and our time segments are too short, thus the error bars are still large. Shorter data intervals will reduce errors, but the power spectrum becomes unreliable. Nevertheless, the QPO PDS still shows clear background and conspicuous peak with larger power at $\sim$ 2.5 Hz, but the non-QPO PDS is noisy and the peak position is random and cannot be identified.

\section{GTI comparison of three instruments}
\begin{figure}
    \begin{minipage}{0.5\textwidth}
		\includegraphics[width=1\textwidth]{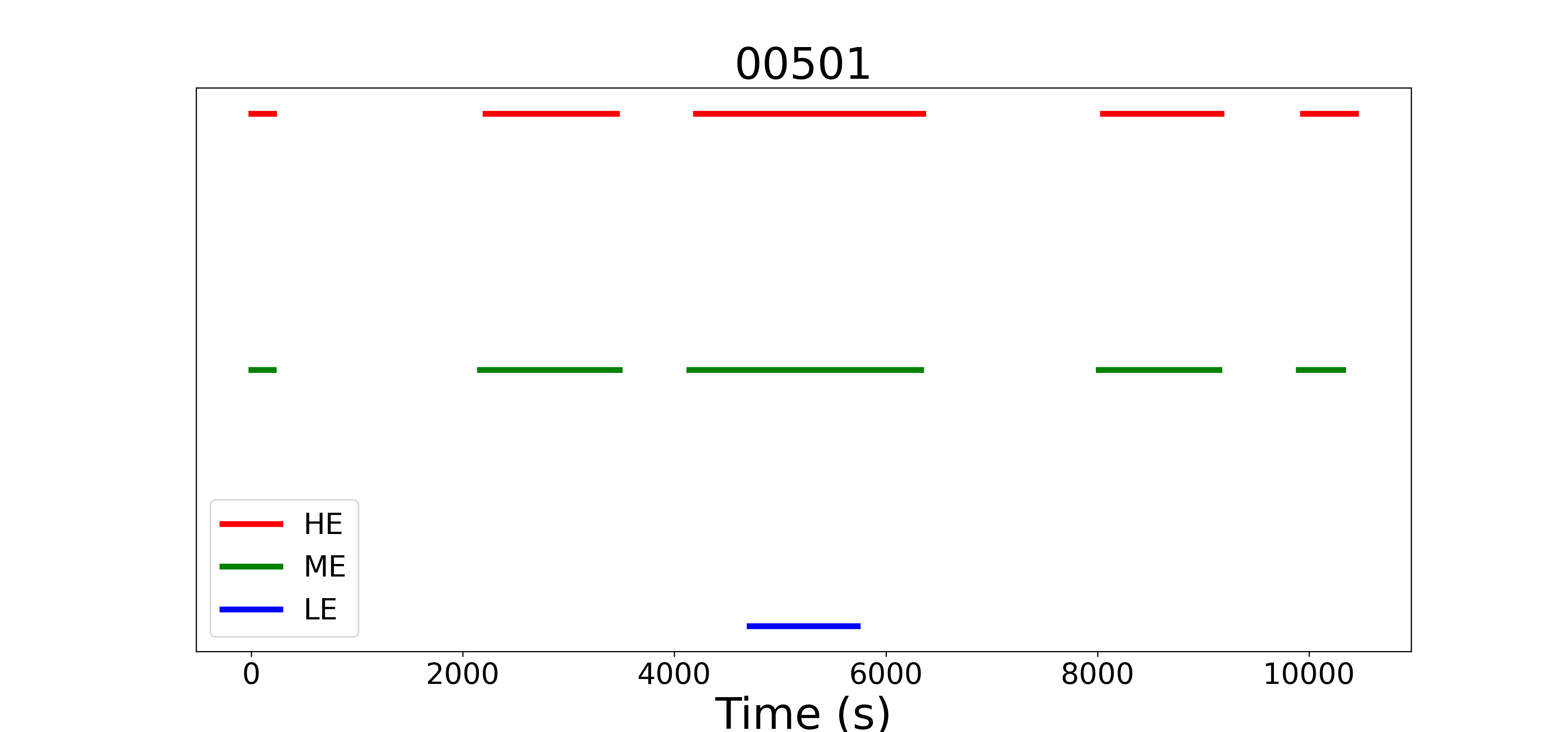}
	\end{minipage}
	\begin{minipage}{0.5\textwidth}
		\includegraphics[width=1\textwidth]{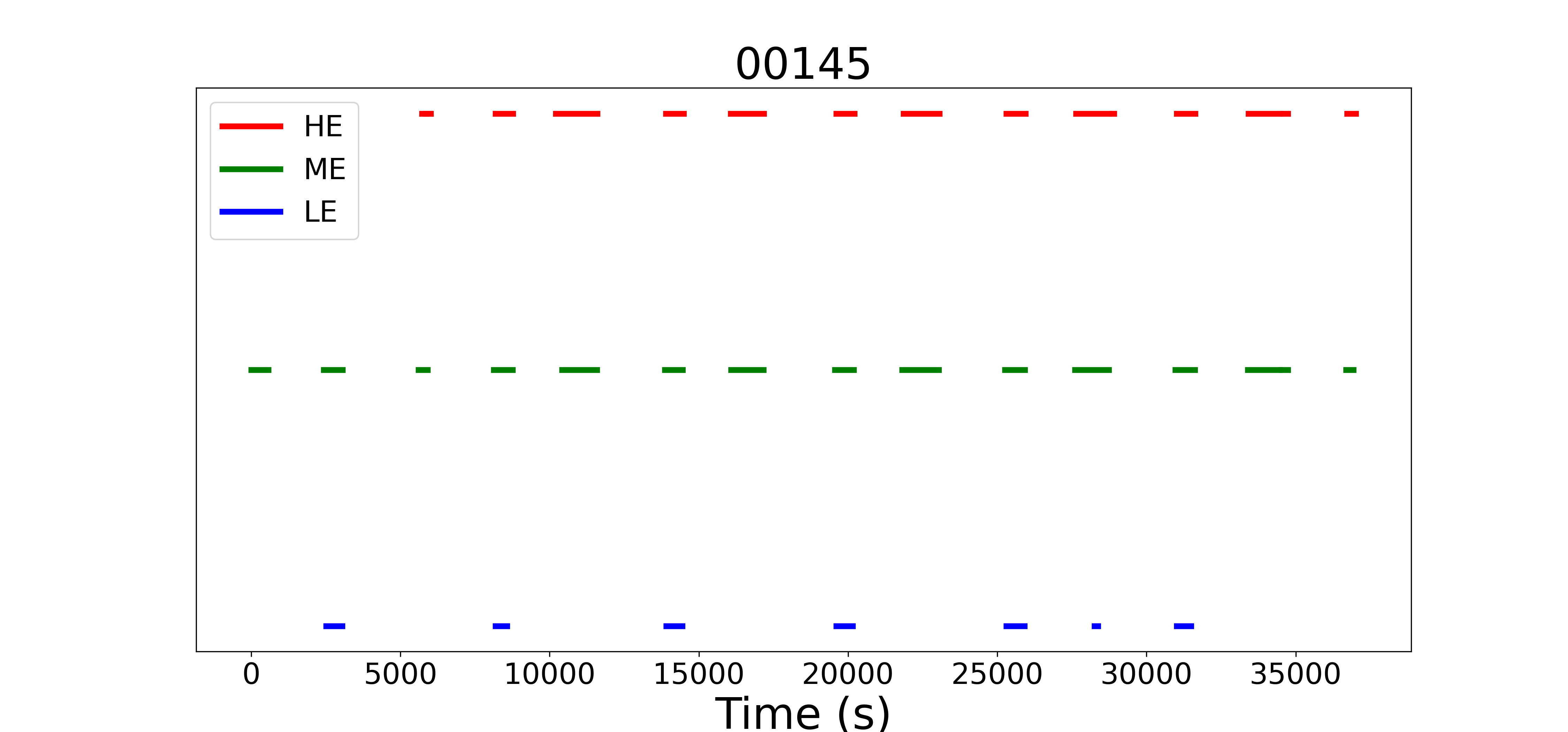}
	\end{minipage}
	\caption{The GTI of LE, ME and HE for Obs. 501 (top) and 145 (bottom).}
	\label{fig:GTI}
\end{figure}

The wavelet analysis is better performed on continuous data, large gaps in time series will reduce accuracy. After checking the GTI of our data, we believe that it is much safer to perform wavelet analysis on time series separated by GTI for each observation. Figure~\ref{fig:GTI} presented the GTI distributions of Obs. 501 and 145. Normally, the HE and ME light curves are separated into $\sim$ 5 time series as the top panel of Figure~\ref{fig:GTI}, while the LE light curve contains 1 or 2 separated time series. The only different observation is Obs. 145, which has a more scattered GTI distribution (see the bottom panel of Figure~\ref{fig:GTI}).

%%%%%%%%%%%%%%%%%%%%%%%%%%%%%%%%%%%%%%%%%%%%%%%%%%

% Don't change these lines
\bsp	% typesetting comment
\label{lastpage}
\end{document}